\documentclass[reqno]{amsart}

\usepackage{amsmath}
\usepackage{amssymb}
\usepackage{amsfonts}
\usepackage[foot]{amsaddr}
\usepackage{mathtools}
\mathtoolsset{%
}

\usepackage[utf8]{inputenc}
\usepackage[T1]{fontenc}

\usepackage[sf,mono=false]{libertine}

\usepackage{dsfont}

\usepackage[%
cal=cm,
]
{mathalfa}

\usepackage[dvipsnames,svgnames]{xcolor}
\colorlet{MyBlue}{DodgerBlue!60!Black}
\colorlet{MyGreen}{DarkGreen!85!Black}

\newcommand{\afterhead}{.}
\usepackage[font=small,labelfont=bf]{caption}
\usepackage{subfigure}
\usepackage{tikz}
\usetikzlibrary{calc,patterns}

\usepackage{acronym}
\usepackage{booktabs}       
\usepackage{latexsym}
\usepackage{paralist}
\usepackage{wasysym}
\usepackage{xspace}

\usepackage[authoryear,sort&compress]{natbib}

\usepackage{hyperref}
\hypersetup{
colorlinks=true,
linktocpage=true,
pdfstartview=FitH,
breaklinks=true,
pdfpagemode=UseNone,
pageanchor=true,
pdfpagemode=UseOutlines,
plainpages=false,
bookmarksnumbered,
bookmarksopen=false,
bookmarksopenlevel=1,
hypertexnames=true,
pdfhighlight=/O,
urlcolor=MyBlue!60!black,linkcolor=MyBlue!70!black,citecolor=DarkGreen!70!black, 
pdftitle={},
pdfauthor={},
pdfsubject={},
pdfkeywords={},
pdfcreator={pdfLaTeX},
pdfproducer={LaTeX with hyperref}
}

\def\EMAIL#1{\email{\href{mailto:#1}{\texttt{\upshape #1}}}}

\numberwithin{equation}{section}  
\usepackage[sort&compress,capitalize,nameinlink]{cleveref}
\crefname{app}{Appendix}{Appendices}

\crefrangeformat{equation}{\upshape(#3#1#4)\textendash(#5#2#6)}

\newcommand{\dd}{\:d}

\newcommand{\eps}{\varepsilon}
\newcommand{\from}{\colon}

\newcommand{\mleq}{\preccurlyeq}

\newcommand{\R}{\mathbb{R}}

\newcommand{\Z}{\mathbb{Z}}
\newcommand{\N}{\mathbb{N}}

\DeclareMathOperator*{\argmin}{arg\,min}

\DeclareMathOperator{\bigoh}{\mathcal{O}}
\newcommand{\smalloh}{o}

\DeclareMathOperator{\ex}{\mathbb{E}}

\DeclareMathOperator{\prob}{\mathbb{P}}

\DeclarePairedDelimiter{\bracks}{[}{]}
\DeclarePairedDelimiter{\parens}{(}{)}

\DeclarePairedDelimiter{\abs}{\lvert}{\rvert}

\DeclarePairedDelimiterX{\braket}[2]{\langle}{\rangle}{#1,#2}

\DeclarePairedDelimiterX{\inner}[2]{\langle}{\rangle}{#1,#2}
\DeclarePairedDelimiterX{\setdef}[2]{\{}{\}}{#1:#2}

\DeclarePairedDelimiterXPP{\probof}[1]{\prob}{(}{)}{}{%

#1}

\DeclarePairedDelimiterXPP{\exof}[1]{\ex}{[}{]}{}{%

#1}

\newcommand{\txs}{\textstyle}
\newcommand{\textpar}[1]{\textup(#1\textup)}

\usepackage[textwidth=30mm,disable]{todonotes}

\newcommand{\debug}[1]{#1}

\theoremstyle{plain}
\newtheorem{theorem}{Theorem}
\newtheorem{corollary}[theorem]{Corollary}
\newtheorem*{corollary*}{Corollary}
\newtheorem{lemma}[theorem]{Lemma}
\newtheorem{proposition}[theorem]{Proposition}

\theoremstyle{definition}
\newtheorem{definition}[theorem]{Definition}
\newtheorem*{definition*}{Definition}

\newtheorem*{hypothesis*}{Hypothesis}

\newcommand{\footnoteOR}[1]{\xspace[#1]}

\newenvironment{Proof}[1][Proof]{\begin{proof}[#1]}{\end{proof}}

\theoremstyle{remark}
\newtheorem{remark}{Remark}
\newtheorem*{remark*}{Remark}
\newtheorem*{notation*}{Notational remark}
\newtheorem{example}{Example}

\numberwithin{theorem}{section}
\numberwithin{remark}{section}
\numberwithin{example}{section}

\newcommand{\game}{\debug \Gamma}

\newcommand{\play}{\debug i}
\newcommand{\playalt}{\debug j}
\newcommand{\nPlayers}{\debug I}
\newcommand{\players}{\mathcal{\debug \nPlayers}}

\newcommand{\graph}{\mathcal{\debug G}}
\newcommand{\vertices}{\mathcal{\debug V}}
\newcommand{\edges}{\mathcal{\debug E}}

\newcommand{\edge}{\debug e}
\newcommand{\edgealt}{\edge'}

\newcommand{\source}{\debug O}

\newcommand{\sink}{\debug D}

\newcommand{\pair}{\debug \play}
\newcommand{\pairalt}{\debug \playalt}

\newcommand{\pairs}{\debug \players}

\newcommand{\rate}{\debug m}
\newcommand{\totrate}{\debug M}
\newcommand{\relrate}{\debug \lambda}

\newcommand{\flow}{\debug f}
\newcommand{\flows}{\mathcal{\debug F}}

\newcommand{\load}{\debug x}
\newcommand{\loads}{\mathcal{\debug X}}

\newcommand{\nRoutes}{\debug P}
\newcommand{\routes}{\mathcal{\debug \nRoutes}}
\newcommand{\route}{\debug p}
\newcommand{\routealt}{\route'}

\newcommand{\simplex}{\Delta}

\newcommand{\ie}{i.e.,\xspace}
\newcommand{\eg}{e.g.,\xspace}

\newcommand{\cost}{\debug c}
\newcommand{\Cost}{\debug C}
\newcommand{\costs}{\mathcal{\debug C}}

\newcommand{\pole}{\debug \omega}
\newcommand{\ind}{\debug \alpha}
\newcommand{\ord}{\debug q}

\newcommand{\test}{\debug g}

\newcommand{\tdegr}{\debug \rho}
\newcommand{\tobj}{\debug G}

\newcommand{\vobj}{\debug V}

\newcommand{\eq}[1]{#1^{\ast}}
\newcommand{\opt}[1]{\tilde#1}
\newcommand{\olim}[1]{\hat#1}

\newcommand{\unitflow}{\debug y}
\newcommand{\unitflows}{\mathcal{\debug Y}}
\newcommand{\unitload}{\debug z}	

\newcommand{\unitfunct}{\debug \zeta}

\newcommand{\pairsalt}{\pairs'}

\newcommand{\fast}[1]{#1_{\textup{fast}}}
\newcommand{\slow}[1]{#1_{\textup{slow}}}
\newcommand{\tight}[1]{#1_{\textup{tight}}}

\newcommand{\obj}{\debug L}

\newcommand{\Val}{W}
\DeclareMathOperator{\Eq}{Eq}
\DeclareMathOperator{\Opt}{Opt}
\DeclareMathOperator{\PoA}{PoA}

\newcommand{\condref}[1]{\hyperref[#1]{Condition~\textpar{\itshape\ref{#1}}}}

\newcommand{\adegr}{\debug a}
\newcommand{\pigourate}{\debug b}
\newcommand{\argmincost}{\routes_{\min}}

\newcommand{\degr}{\debug d}

\newcommand{\gdegr}{\debug g}

\newcommand{\minc}{\debug H}

\newcommand{\pigoualt}{\debug \theta}
\newcommand{\pigoualtlight}{\debug \rho}

\newcommand{\pigouratelight}{\debug b}
\newcommand{\pol}{\debug k}
\newcommand{\run}{\debug n}
\newcommand{\firstconst}{\debug K_{1}}

\newcommand{\secondconst}{\debug K_{\adegr}}

\newcommand{\sumB}{\debug D}
\newcommand{\sumc}{\debug G}
\newcommand{\sumcalt}{\debug B}

\newcommand{\acdef}[1]{\textit{\acl{#1}} \textpar{\acs{#1}}\acused{#1}}

\newacro{NE}{Nash equilibrium}
\newacroplural{NE}[NE]{Nash equilibria}
\newacro{WE}{Wardrop equilibrium}
\newacroplural{WE}[WE]{Wardrop equilibria}
\newacro{SO}{socially optimum}

\newacro{KKT}{Karush\textendash Kuhn\textendash Tucker}
\newacro{OD}[O/D]{origin-destination}
\newacro{PoA}{price of anarchy}
\newacro{BPR}{Bureau of Public Roads}

\begin{document}


\title
[When is selfish routing bad?]
{When is selfish routing bad?\\
The price of anarchy in light and heavy traffic}

\author
[R.~Colini-Baldeschi]
{Riccardo Colini-Baldeschi$^{\ast}$}
\address{$^{\ast}$ Core Data Science Group, Facebook Inc., 1 Rathbone Place, London, W1T 1FB, UK.}
\EMAIL{rickuz@fb.com}

\author
[R.~Cominetti]
{Roberto Cominetti$^{\ddag}$}
\address{$^{\ddag}$ Facultad de Ingenier\'ia y Ciencias, Universidad Adolfo Ib\'a\~nez, Santiago, Chile.}
\EMAIL{roberto.cominetti@uai.cl}

\author
[P.~Mertikopoulos]
{Panayotis Mertikopoulos$^{\S}$}
\address{$^{\S}$ Univ. Grenoble Alpes, CNRS, Inria, LIG, F-38000 Grenoble, France.}
\EMAIL{panayotis.mertikopoulos@imag.fr}

\author
[M.~Scarsini]
{Marco Scarsini$^{\P}$}
\address{$^{\P}$ Dipartimento di Economia e Finanza, LUISS, Viale Romania 32, 00197 Roma, Italy.}
\EMAIL{marco.scarsini@luiss.it}

%
%
\thanks{We thank an anonymous reviewer of an earlier conference version of this paper for suggesting one of the examples with variable inflows in \cref{sec:variable}.}
\thanks{This research benefited from the support of the FMJH Program PGMO under grant HEAVY.NET and from the support of EDF, Thales, and Orange.
R.~Colini-Baldeschi and M.~Scarsini are members of GNAMPA-INdAM.
R.~Cominetti and P.~Mertikopoulos gratefully acknowledge the support and hospitality of LUISS during a visit in which this research was initiated.
R.~Cominetti's research is also supported by FONDECYT 1130564 and N\'ucleo Milenio ICM/FIC RC130003 ``\emph{Informaci\'on y Coordinaci\'on en Redes}.''
P.~Mertikopoulos was partially supported by
the ECOS/CONICYT Grant C15E03
and the Huawei HIRP Flagship project ULTRON.
P.~Mertikopoulos and M.~Scarsini also gratefully acknowledge the support and hospitality of FONDECYT 1130564 and N\'ucleo Milenio ``\emph{Informaci\'on y Coordinaci\'on en Redes}.''}


\begin{abstract}
%
%
This paper examines the behavior of the \acl{PoA} as a function of the traffic inflow in nonatomic congestion games with multiple \ac{OD} pairs.
Empirical studies in real-world networks show that the \acl{PoA} is close to $1$ in both light and heavy traffic, thus raising the question:
can these observations be justified theoretically?
We first show that this is not always the case:
the \acl{PoA} may remain a positive distance away from $1$ for all values of the traffic inflow, even in simple three-link networks with a single \ac{OD} pair and smooth, convex costs.
On the other hand, for a large class of cost functions (including all polynomials), the \acl{PoA} \emph{does} converge to $1$ in both heavy and light traffic, irrespective of the network topology and the number of \ac{OD} pairs in the network.
We also examine the rate of convergence of the \acl{PoA}, and we show that it follows a power law whose degree can be computed explicitly when the network's cost functions are polynomials.
\end{abstract}

\allowdisplaybreaks
\maketitle
\acresetall

\begin{center}
\small
``\emph{Traffic congestion is caused by vehicles, not by people in themselves.}''
\medskip

\begin{flushright}
\textemdash\ Jane Jacobs, \emph{The Death and Life of Great American Cities}
\end{flushright}
\end{center}

\section{Introduction}
\label{sec:introduction}

Almost every commuter in a major metropolitan area has experienced the frustration of being stuck in traffic.
At best, this might mean being late for dinner;
at worst, it means more accidents and altercations, not to mention the vastly increased damage to the environment caused by huge numbers of idling engines.

To name but an infamous example, China's G110 traffic jam in August 2010 brought to a standstill thousands of vehicles for 100 kilometers between Hebei and Inner Mongolia.
The snarl-up lasted twelve days and resulted in drivers being unable to move for more than 1 kilometer per day, reportedly spending up to five days trapped in the jam.
Not caused by weather or a natural disaster, this massive $10$-day tie-up was instead laid at the feet of a bevy of trucks swarming on the shortest route to Beijing, thus clogging the highway to a halt (while ironically carrying supplies for construction work to ease congestion).
This, therefore, raises the question:
\emph{how much better would things have been if all traffic had been routed by a social planner who could calculate 
\textpar{and enforce} the optimum traffic assignment?}

In game-theoretic terms, this question boils down to the inefficiency of \aclp{NE} that are not Pareto optimal.
The most widely used quantitative measure of this inefficiency is the so-called \acdef{PoA}:
introduced by \citet{KouPap:STACS1999} and so dubbed by \citet{Pap:PACM2001}, the \acl{PoA} is simply the ratio of the social cost of the least efficient \acl{NE} divided by the minimum achievable social cost.
By virtue of this straightforward definition, deriving worst-case bounds for the \acl{PoA} has given rise to a vigorous literature at the interface of operations research, economics and computer science, often leading to surprising and counter-intuitive results.

In the context of network congestion, \cite{Pig:Macmillan1920} was probably the first to note the inefficiency of selfish routing, and his elementary two-road example with a \ac{PoA} of $4/3$ is one of the two prototypical examples thereof.
The other example is due to \cite{Bra:U1968}, and consists of a four-road network where the addition of a zero-cost segment makes things just as bad as in the Pigou case.
These two examples were the starting point for \cite{RouTar:JACM2002} who showed that the \acl{PoA} in (nonatomic) routing games with affine costs may not exceed $4/3$.
On the other hand, if the network's cost functions are polynomials of degree at most $d$, the \acl{PoA} may become as high as $\Theta(d/\log d)$, implying that selfish routing can become arbitrarily bad in networks with polynomial costs \citep{Rou:JCSS2003}.

By this token, and given the typically nonlinear relation between traffic loads and travel times, the intervention of a central planner seems necessary in order to regain some degree of efficiency.
At the same time however, these worst-case instances are typically realized in networks with delicately tuned traffic loads and costs:
if a network operates beyond this regime, it is not clear whether the \acl{PoA} remains high.
In view of this, our aim in this paper is to examine the asymptotic behavior of the \acl{PoA} at both ends of the congestion spectrum:
\emph{light and heavy traffic}.

Using both analytical and numerical methods, a very recent study by \cite{OHaConWat:TRB2016} suggests that the \acl{PoA} is usually close to $1$ for both high and low traffic, and it fluctuates in the intermediate regime (typically exhibiting multiple local maxima).
In a similar setting, \cite{MonBenPil:ArXiv2017} used a huge dataset on commuting students in Singapore to estimate the so-called ``stress of catastrophe'':
this majorant of the ordinary \acl{PoA} was estimated to a value between $1.11$ and $1.22$, suggesting that the actual value of the \acl{PoA} in Singapore is lower (and definitely below the $4/3$ worst-case bound of the Pigou/Braess examples).

\smallskip

All this leads to the following natural questions:
\begin{enumerate}
[\indent\itshape a\upshape)]

\item
Under what conditions does the \acl{PoA} converge to $1$ in light/heavy traffic?

\item
Do these conditions depend on the network topology, its cost functions, or both?

\item
Can general results be obtained for networks with multiple \ac{OD} pairs?

\item
When these conditions are satisfied, how fast is this convergence?
\end{enumerate}


\subsection{Our contributions}
\label{sec:results}

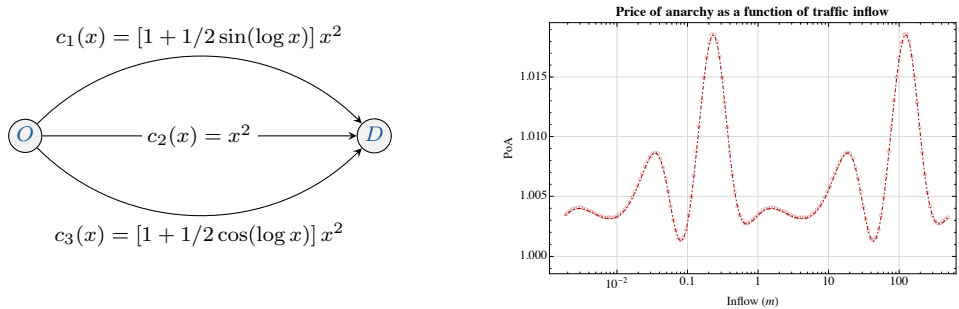
\begin{figure}[t]
\centering
\footnotesize
\subfigure{

\begin{tikzpicture}
[scale=1.1,
nodestyle/.style={circle,draw=black,fill=gray!10, inner sep=1pt, text = MyBlue},
edgestyle/.style={-},
>=stealth]

\coordinate (A) at (-\textwidth/6,0);
\coordinate (B) at (\textwidth/6,0);

\node (A) at (A) [nodestyle] {$\source\vphantom{bp}$};
\node (B) at (B) [nodestyle] {$\sink\vphantom{bp}$};
\node (phantom) at (0,-2) {};

\draw [edgestyle,->] (A) to [bend left = 45] node [midway, above] {$\cost_{1}(\load) = \bracks*{1 + 1/2 \sin(\log\load)} \, \load^{2}$} (B);
\draw [edgestyle,->] (A) to node [midway, fill = white, inner sep = 2pt] {$\cost_{2}(\load) = \load^{2}$} (B);
\draw [edgestyle,->] (A) to [bend right = 45] node [midway, below] {$\cost_{3}(\load) = \bracks*{1 + 1/2 \cos(\log\load)} \, \load^{2}$} (B);

\end{tikzpicture}}
\hfill
\subfigure{\includegraphics[width=.48\textwidth,]{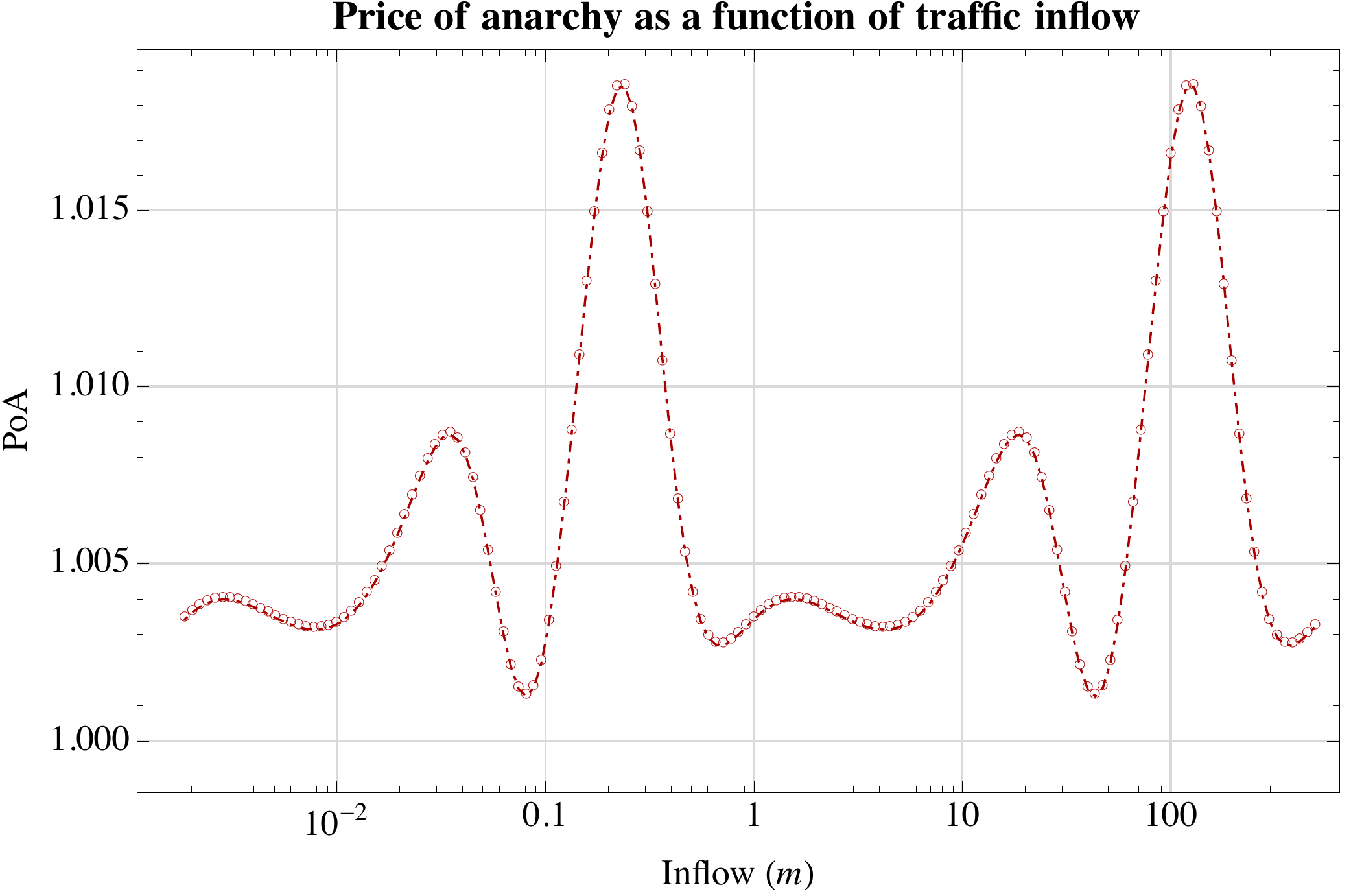}}
\caption{A network where selfish routing remains inefficient for both light and heavy traffic.}
\vspace{-3ex}
\label{fig:parallel}
\end{figure}

Our first result is a cautionary tale:
we show that the \acl{PoA} may oscillate between two bounds strictly greater than $1$ for all values of the traffic inflow, even in simple parallel-link networks with a single \ac{OD} pair (cf.~\cref{fig:parallel}).
The cost functions in our example are convex and differentiable, so neither convexity nor smoothness seem to play a major role in the efficiency of selfish routing.
Moreover, our construction only involves a three-link network, so such phenomena may arise in any network containing such a three-link component.


Heuristically, the reason for this irregular \textendash\ and, perhaps, counter-intuitive \textendash\ behavior is that the growth rate of the network's cost functions exhibits higher-order oscillations which persist at any scale, in both light and heavy traffic.
To dispense with such pathologies, we focus on networks whose cost functions $\cost_\edge(\load)$ are asymptotically comparable to a \emph{benchmark function} $\cost(\load)$ which is itself assumed to be \emph{regularly varying} (cf.~\cref{def:regvar}).
In so doing, we obtain a classification of the network's edges, paths, and \ac{OD} pairs as \emph{fast}, \emph{slow} or \emph{tight} relative to the chosen benchmark.
Then, thanks to this classification, we obtain the following general result:
\emph{If the routing cost of the ``most costly'' \ac{OD} pair in the network behaves like the benchmark, the network's \acl{PoA} converges to $1$ in both light and heavy traffic.}
%

Polynomial cost functions satisfy all of the above requirements, leading to the comprehensive asymptotic principle:
\smallskip
\begin{quotation}
\centering
\itshape
In networks with polynomial costs,\\
the \acl{PoA} becomes $1$ under both light and heavy traffic.
\end{quotation}
\smallskip
In other words, a benevolent social planner with full control of traffic assignment would not do any better than selfish agents in conditions of high or low congestion.
In particular, only if the traffic falls in an intermediate regime can there be a substantial gap between optimum and equilibrium states.

To assess how wide this intermediate regime might be in practice, we also examine the speed at which the \acl{PoA} converges to $1$ as a function of the traffic inflow.
Specializing to networks with polynomial costs, we find that in both regimes 
the convergence follows a power law with respect to the total traffic inflow, and we derive explicit sharp estimates for the corresponding rates.

\subsection{Related work}
\label{sec:related}

Establishing worst-case bounds for the \acl{PoA} under different conditions has been a staple of the literature on congestion games ever since the seminal result of \cite{RouTar:JACM2002}.
In words, this result states that in networks with affine costs the \acl{PoA} is no higher than $4/3$, independently of the network topology and/or the number of \ac{OD} pairs in the network.
Furthermore, this bound is sharp in that, for every $\totrate>0$, there exists a network with traffic inflow $\totrate$ and affine costs such that the \acl{PoA} is \emph{exactly} equal to $4/3$.
Importantly, our analysis shows that the order of the quantifiers in the above statement \emph{cannot be exchanged:}
in any network with affine costs, the \acl{PoA} gets arbitrarily close to $1$ if the traffic inflow is sufficiently large or small.

Worst-case bounds for the \acl{PoA} have been obtained for larger classes of cost functions.
For polynomial costs with degree at most $\degr$, \cite{Rou:JCSS2003} showed that the worst possible instance grows as $\Theta(\degr/\log\degr)$ while \cite{DumGai:INE2006} provided sharper bounds for monomials of maximum degree $\degr$ and minimum degree $\ord$.
Extending the above results, \cite{RouTar:GEB2004} provided a unifying result for costs that are differentiable with $\load\cost(\load)$ convex, while \cite{CorSchSti:MOR2004,CorSchSti:GEB2008} considered less regular classes of cost functions.
\cite{CorSchSti:OR2007} also studied the \acl{PoA} when the goal is to minimize the maximum \textendash\ rather than the average \textendash\ latency in the network.
For a survey, the reader is referred to \cite{Rou:AGT2007}.

In a more practical setting, \cite{YouGasJeo:PRL2008} studied the difference between optimal and actual system performance in real transportation networks, focusing in particular on Boston's road network.
They observed that the \acl{PoA} depends crucially on the total traffic inflow:
it starts at $1$, it then grows with some oscillations, and ultimately returns to $1$ as the flow increases.
\cite{GonGraAndLyg:IEEECDC2015} studied optimal scheduling for the electricity demand of a fleet of plug-in electric vehicles:
without using the term, they showed that the \acl{PoA} goes to $1$ as the number of vehicles grows.
\cite{ColTao:EC2016} showed that in large Walrasian auctions and in large Fisher markets the \acl{PoA} goes to one as the market size increases.
Finally, \cite{FelImmLucRouSyr:STOC2016} took a different asymptotic approach and considered atomic games where the number of players grows to infinity.
Applying the notion of $(\lambda,\mu)$-smoothness to the resulting sequence of atomic games, they showed that the \acl{PoA} converges to the corresponding nonatomic limit.

From an analytic standpoint, the closest antecedent to our paper is the recent work of \cite{ColComSca:SAGT2016} who 
studied the heavy traffic limit of the \acl{PoA} in parallel networks with a single \ac{OD} pair.
Their analysis identified that regular variation plays an important role in this setting;
however, it offered no insights into non-parallel networks with multiple \ac{OD} pairs or the light traffic regime.
Our paper provides an in-depth answer to these questions:
we show that
\begin{inparaenum}
[(\itshape a\upshape)]
\item
regular variation yields asymptotic efficiency under both light and heavy traffic conditions;
\item
the topology of the network doesn't matter;
and
\item
the existence of several \ac{OD} pairs doesn't matter as long as they admit a common benchmark (which is always the case if the network's cost functions are polynomial).
\end{inparaenum}

Building on a previous unpublished version of the present paper, \citet{WuMohChe:ArXiv2017} introduced a class of congestion games, called \emph{scalable}, whose \acl{PoA} converges to $1$ as the total demand diverges.
They also computed the rate of convergence of the \acl{PoA} for the special case of \acs{BPR} cost functions of the same degree.
\cite{Sti:Springer2014} also studied the behavior of the \acl{PoA} for queueing networks in heavy traffic under various assumptions on the structure of the network and on the stochastic properties of the queues.
His results can be used for the analysis of network routing models with capacity constraints.

Our work should also be compared to that of \cite{MonBenPil:ArXiv2017} who performed an empirical study of the \acl{PoA} based on data from thousands of commuting students in Singapore.
Focusing on the network's ``stress of catastrophe'' (an empirical majorant of the network's \acl{PoA}), they showed that routing choices are near-optimal and the incurred \acl{PoA} is much lower than what traditional worst-case bounds suggest.
Interestingly, the study of \cite{MonBenPil:ArXiv2017} also suggests that the Singapore road network is often lightly congested:
as such, their results can be seen as a practical validation of the light traffic results presented here (and, conversely, our results provide a theoretical justification for their empirical observations).


\subsection{Outline of the paper}
\label{sec:outline}

The paper is organized as follows.
In \cref{sec:model}, we introduce the basic model and concepts that will be used in the rest of the paper.
\cref{sec:example} provides two motivating examples for the analysis to follow.
In \cref{sec:single}, we treat networks with a single \ac{OD} pair, whereas \cref{sec:multi} examines networks with multiple \ac{OD} pairs. 
The more complicated case of variable relative inflows is treated  in \cref{sec:variable}. 
Finally, in \cref{sec:rate}, we study the rate of convergence of the \acl{PoA} in light and heavy traffic.
To streamline our presentation, the proofs of our main results have been relegated to a series of appendices at the end of the paper.

\section{Model and preliminaries}
\label{sec:model}


\subsection{Network model}
\label{sec:network}

Following \cite{BecMcGWin:Yale1956} and \cite{RouTar:JACM2002}, the basic component of our model will be a directed multi-graph $\graph\equiv\graph(\vertices,\edges)$ with vertex set $\vertices$ and edge set $\edges$ (both finite).
We further assume that there is a finite set
of \acdef{OD} pairs indexed by $\pair\in\pairs$, each with an individual \emph{traffic demand} $\rate^{\pair}\geq 0$ that is to be routed from the pair's \emph{origin node} $\source^{\pair}\in\vertices$ to its \emph{destination} $\sink^{\pair}\in\vertices$.
To route this traffic, the $\pair$-th \ac{OD} pair employs a set $\routes^{\pair}$ of \emph{paths} joining $\source^{\pair}$ to $\sink^{\pair}$, with each path $\route\in\routes^{\pair}$ comprising a sequence of edges that meet head-to-tail in the usual way;
specifically, we do not assume that $\routes^{\pair}$ is necessarily the set of \emph{all} paths joining $\source^{\pair}$ to $\sink^{\pair}$, but only some subset thereof.%
\footnoteOR{This distinction is particularly relevant for packet-switched networks (such as the Internet) where only paths with a low hop count are typically employed.}
For bookkeeping reasons, we will also make the following standing assumptions throughout our paper:
\vspace{.5ex}
\begin{enumerate}
[\indent\itshape a\upshape)]
\addtolength{\itemsep}{.5ex}
\item
The \emph{total inflow rate} $\totrate = \sum_{\pair\in\pairs} \rate^{\pair}$ is positive (so there is a nonzero amount of traffic in the network).
\item
The path sets $\routes^{\pair}$ are disjoint (which in particular holds trivially if all pairs $(\source^{\pair},\sink^{\pair})$ are distinct).
\end{enumerate}
\vspace{.5ex}

Now, writing $\routes \equiv \bigcup_{\pair\in\pairs} \routes^{\pair}$ for the union of all such paths, the set of feasible \emph{routing flows} $\flow = (\flow_{\route})_{\route\in\routes}$ in the network is defined as
\begin{equation}
\label{eq:flows}
\txs
\flows
	= \setdef*{\flow\in\R_{+}^{\routes}}{\text{$\sum_{\route\in\routes^{\pair}} \flow_{\route} = \rate^{\pair}$ for all $\pair\in\pairs$}}.
\end{equation}
In turn, a routing flow $\flow\in\flows$ induces a \emph{load} on each edge $\edge\in\edges$ as
\begin{equation}
\label{eq:load}
\load_{\edge}
	= \sum_{\route\ni\edge} \flow_{\route},
\end{equation}
and we write $\load = (\load_{\edge})_{\edge\in\edges}$ for the corresponding \emph{load profile} on the network. 
Given all this, the delay (or latency) experienced by an infinitesimal traffic element traversing edge $\edge$ is determined by a nondecreasing continuous \emph{cost function} $\cost_{\edge}\from[0,\infty)\to[0,\infty)$:
more precisely, if $\load = (\load_{\edge})_{\edge\in\edges}$ is the load profile induced by a feasible routing flow $\flow = (\flow_{\route})_{\route\in\routes}$, the incurred delay on edge $\edge\in\edges$ is $\cost_{\edge}(\load_{\edge})$.
Hence, with a slight abuse of notation, the associated cost of path $\route\in\routes$ will be given by
\begin{equation}
\label{eq:cost-path}
\cost_{\route}(\flow)
	= \sum_{\edge\in\route} \cost_{\edge}(\load_{\edge}).
\end{equation}

Putting together all of the above, the tuple $\game = (\graph,\pairs,\{\rate^{\pair}\}_{\pair\in\pairs},\{\routes^{\pair}\}_{\pair\in\pairs},\{\cost_{\edge}\}_{\edge\in\edges})$ will be referred to as a \emph{\textpar{nonatomic} routing game}.
When we want to explicitly keep track of the total inflow rate $\totrate = \sum_{\pair}\rate^{\pair}$, we write $\game_{\totrate}$ instead of $\game$;
also, when there is a single \ac{OD} pair, we will drop all reference to $\pair$ and $\pairs$ altogether.

\subsection{Equilibrium, optimality, and the \acl{PoA}}
\label{sec:PoA}

In routing games, the notion of \acl{NE} is captured by Wardrop's first principle:
\emph{at equilibrium, the delays along utilized paths are equal and no higher than those that would be experienced by an infinitesimal traffic element going through an unused route} \citep{War:PICE1952}.

Formally, a routing flow $\eq{\flow}$ is said to be a \acdef{WE} of $\game$ if, for all $\pair\in\pairs$, we have
\begin{equation}
\label{eq:Wardrop}
\cost_{\route}(\eq{\flow})
	\leq \cost_{\routealt}(\eq{\flow})
	\quad
	\text{for all $\route,\routealt\in\routes^{\pair}$ such that $\eq{\flow}_{\route}>0$}.
\end{equation}
From the work of \cite{BecMcGWin:Yale1956}, it is known that \aclp{WE} coincide with the solutions of the (convex) minimization problem
\begin{equation}
\label{eq:WE}
\tag{WE}
\begin{aligned}
\textrm{minimize}
	&\quad
	\sum_{\edge\in\edges} \Cost_{\edge}(\load_{\edge}),
	\\
\textrm{subject to}
	&\quad
	\load_{\edge} = \sum_{\route\ni\edge} \flow_{\route},
	\;
	\flow\in\flows,
\end{aligned}
\end{equation}
where
$\Cost_{\edge}(\load_{\edge}) = \int_{0}^{\load_{\edge}} \cost_{\edge}(w) \dd w$ denotes the primitive of $\cost_{\edge}$.
Analogously, \acdef{SO} flows are defined as solutions of the total cost minimization problem
\begin{equation}
\label{eq:SO}
\tag{SO}
\begin{aligned}
\textrm{minimize}
	&\quad
	\sum_{\route\in\routes} \flow_{\route} \cost_{\route}(\flow_{\route}),
	\\
\textrm{subject to}
	&\quad
	\flow\in\flows.
\end{aligned}
\end{equation}

By a simple rearrangement of terms, the objective function of \eqref{eq:SO} can be rewritten as $\obj(\load) = \sum_{\edge\in\edges} \load_{\edge} \cost_{\edge}(\load_{\edge})$, so the value of the above problem can be expressed equivalently (and somewhat more concisely) as
\begin{equation}
\Opt(\game)
	= \min_{\load\in\loads} \obj(\load),
\end{equation}
where $\loads = \setdef{x\in\R_{+}^{\edges}}{\load_{\edge} = \sum_{\route\ni\edge} \flow_{\route}, \flow\in\flows}$ denotes the set of all load profiles of the form \eqref{eq:load}.
Thus, to quantify the gap between solutions to \eqref{eq:WE} and \eqref{eq:SO}, let
\begin{equation}
\label{eq:values}
\Eq(\game)
	= \obj(\eq\load)
\end{equation}
where $\eq\load$ is the load profile induced by a \acl{WE} $\eq\flow$ of $\game$ (by a standard result of \cite{BecMcGWin:Yale1956}, all such flows incur the same total cost).
The game's \acdef{PoA} is then defined as
\begin{equation}
\label{eq:PoA}
\PoA(\game)
	= \frac{\Eq(\game)}{\Opt(\game)}.
\end{equation}
For this ratio to be well-defined, we must have $\Opt(\game)>0$;
otherwise, if this is not the case, we will vacuously set $\PoA(\game) = 1$.
To avoid such technicalities, we will tacitly assume that $\Opt(\game)>0$ throughout.

Of course, $\PoA(\game) \geq 1$ with equality if and only if \aclp{WE} are also socially efficient.
Our main objective in what follows will be to study the asymptotic behavior of this ratio when $\totrate\to0$ or $\totrate\to\infty$.

\section{First results}
\label{sec:example}


\subsection{Sioux Falls: a representative case study}
\label{sec:Sioux-Falls}

\begin{figure}[t]
\centering
\footnotesize
\subfigure[The Sioux Falls road network.]{%
\begin{tikzpicture}
\node (graph) at (0,0) {\includegraphics[width=.36\textwidth]{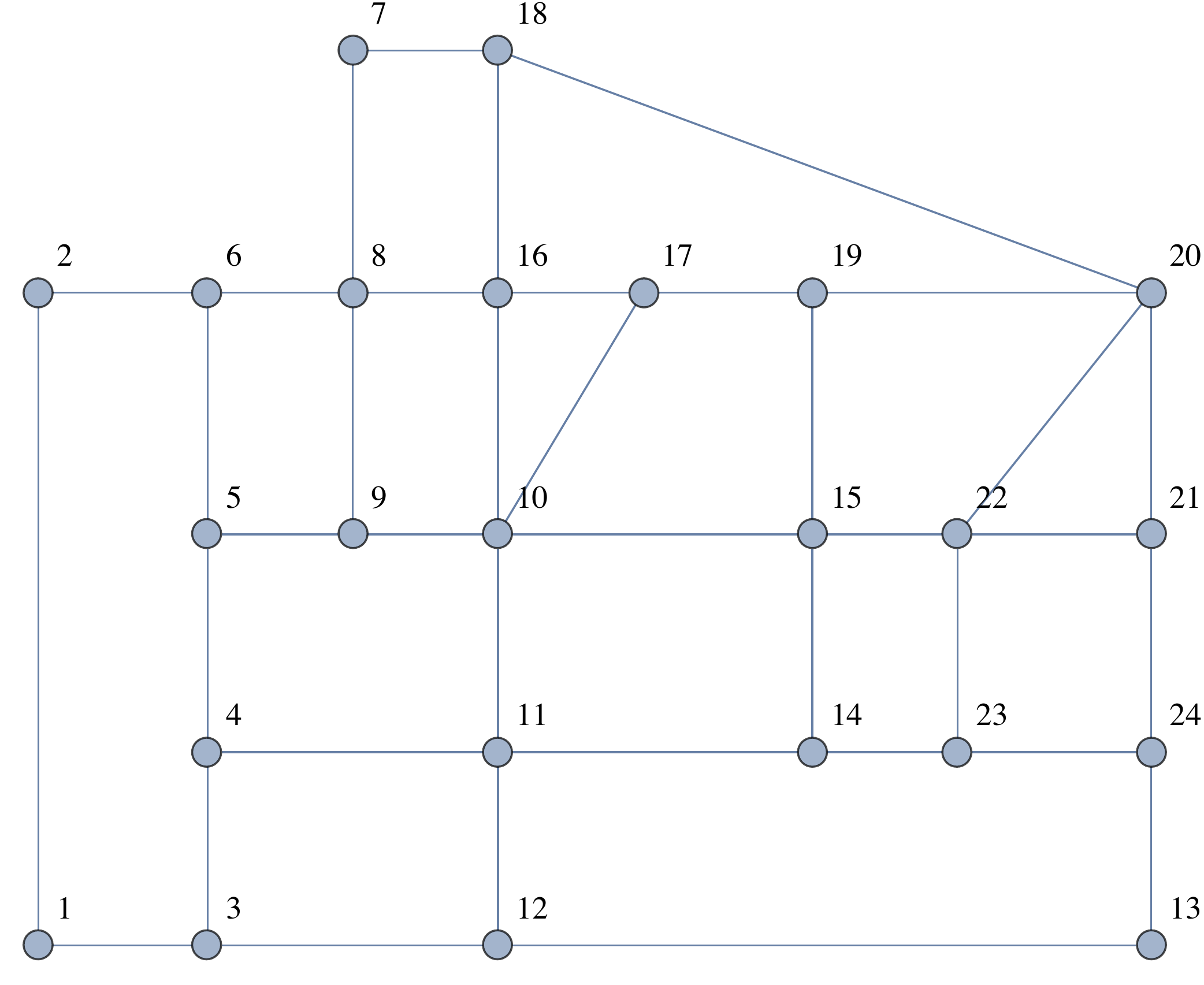}};
\node (phantom) at (0,-2.3) {};
\end{tikzpicture}
\label{fig:SiouxFalls-network}}
\hfill
\subfigure[Anarchy and efficiency in Sioux Falls.]{%
\includegraphics[width=.5\textwidth]{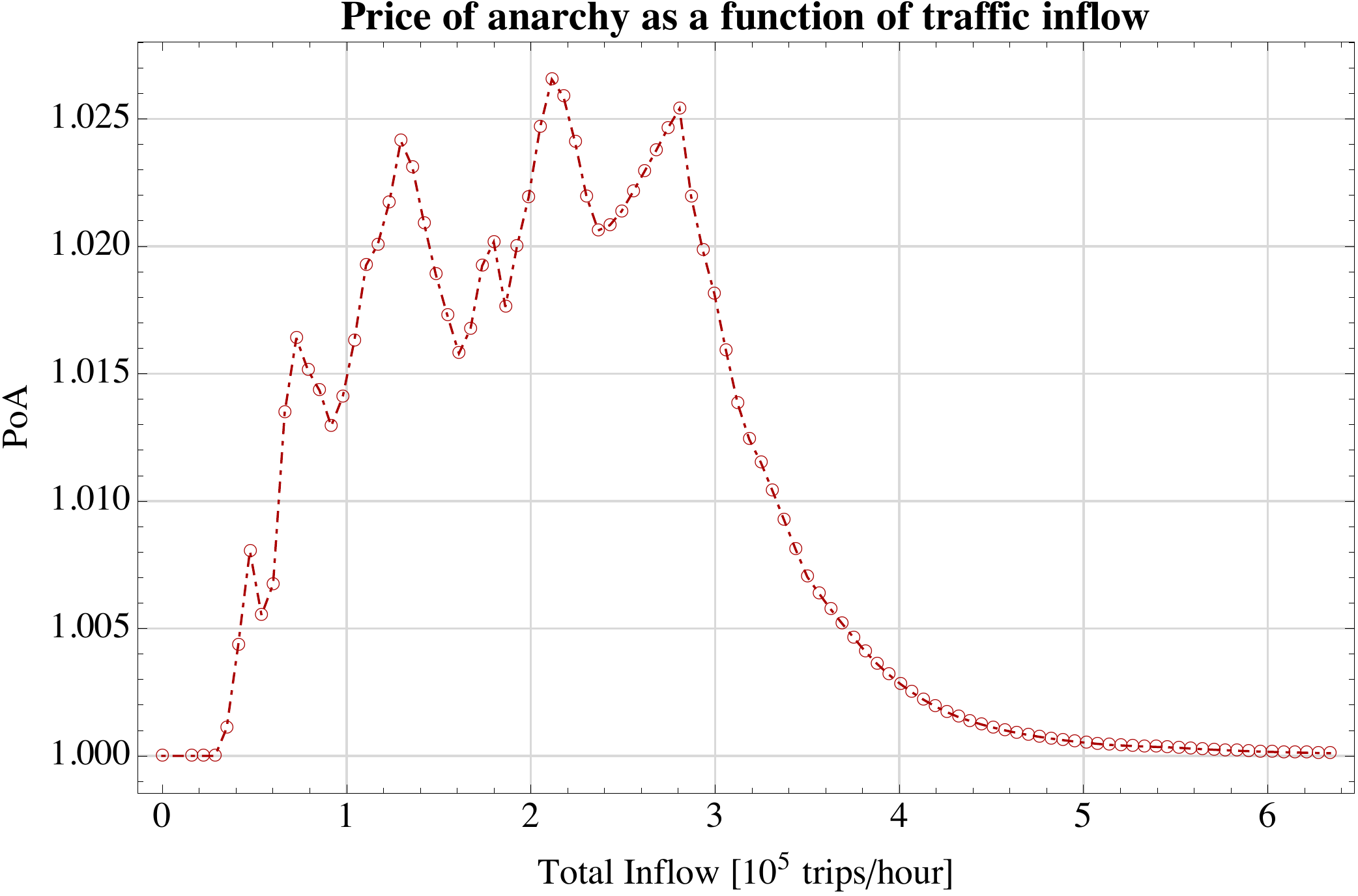}
\label{fig:SiouxFalls-PoA}}
\caption{The \acl{PoA} in the Sioux Falls metropolitan area as a function of the traffic inflow.}
\vspace{-1em}
\label{fig:SiouxFalls}
\end{figure}

To motivate our analysis, we begin by examining the behavior of the \acl{PoA} in the road network of Sioux Falls, a standard case study in the transportation literature.
For concreteness, the network's (two-way) arterial roads are shown in \cref{fig:SiouxFalls-network} and their delay functions are taken to be of the \acs{BPR} (\acl{BPR}) type
\begin{equation}
\label{eq:BPR}
\cost_{\edge}(\load)
	= a_{\edge} + b_{\edge} \load^{\degr_\edge}
\end{equation}
with coefficients $a_{\edge}, b_{\edge}$ and degrees $d_{\edge}$ (typically $d_{\edge}=4$) taken from the standard reference work of \citet[Table~1]{LMP75}.
To analyze the network's \acl{PoA} as a function of the total traffic inflow, we considered all $528$ \ac{OD} pairs with nonzero inflow, and for each \ac{OD} pair $\pair\in\pairs$, we restricted $\routes_{\pair}$ to contain only the five shortest paths in terms of free-flow travel time (\ie the time taken to traverse a path when empty).
We then scaled up or down these inflows preserving the ratios between different \ac{OD} pairs and we plotted the network's \acl{PoA} for various values of the total inflow $\totrate$.


As can be seen from \cref{fig:SiouxFalls-PoA}, the network's \acl{PoA} is identically equal to $1$ when the total inflow is small enough (approximately up to $3.7\times10^{4}$ trips per hour);
for intermediate values of $\totrate$, the \acl{PoA} becomes strictly greater than $1$,
and, ultimately, it decreases monotonically to $1$ in the heavy traffic limit.
Interestingly, \cite{LMP75} report a value of $\totrate_{\textrm{avg}} \approx 3.6\times 10^{5}\,\textrm{trips/hour}$ for the network's median traffic inflow;
this value is well within the range where the \acl{PoA} decreases monotonically to $1$ and, indeed, the observed value is approximately equal to $1.005$, indicating a $0.5\%$ difference between socially optimum and equilibrium flows under median traffic conditions.

Similar conclusions have been drawn in the literature from empirical studies in London, New York and Boston \citep{YouGasJeo:PRL2008}, as well as Sioux Falls with different subsets of \ac{OD} pairs and connecting paths per pair \citep{OHaConWat:TRB2016}.
In particular, in all cases, it was observed that there is an initial interval of values of $\totrate$ for which the \acl{PoA} is identically equal to $1$;
our first result shows that this is not mere happenstance:

\begin{proposition}
\label{prop:BPR-light}
If the network's cost functions are of the form \eqref{eq:BPR} with $a_{\edge}, b_{\edge}>0$ and $\degr_{\edge}\equiv\degr$ for all $\edge\in\edges$, we have $\PoA(\game_{\totrate})=1$  for all sufficiently small $\totrate$. 
\end{proposition}

In fact, as the following result shows, this behavior arises whenever each \ac{OD} pair admits a single ``best'' path under zero inflow:

\begin{proposition}
\label{prop:PoA-equal-one}
Let $\argmincost^{\pair}=\argmin_{\route\in\routes^{\pair}}\cost_{\route}(0)$ denote the set of minimal cost paths of the $\pair$-th \ac{OD} pair under zero inflow.
If $\argmincost^{\pair}$ is a singleton for all $\pair\in\pairs$, we have $\PoA(\game_{\totrate})=1$ for all sufficiently small $\totrate$.
\end{proposition}


The above results (both proven in \cref{app:example}) provide a reasonable theoretical justification for the light traffic behavior of the \acl{PoA} that is observed in \cref{fig:SiouxFalls} (the heavy traffic limit is discussed in detail in the next sections).
At the same time however, the \acs{BPR} and ``unique best path'' assumptions in \cref{prop:BPR-light,prop:PoA-equal-one} respectively suggest that there is a finer mechanism at play which becomes apparent when the total cost at low traffic depends more delicately on the distribution of traffic in the network.
We make this precise in the following section where we provide an example of a three-link network where the \acl{PoA} oscillates between two values strictly greater than $1$, for all values of the traffic inflow.

\subsection{A network where selfish routing is always inefficient}
\label{sec:counterexample}

To construct an example of an ``always inefficient'' network, our approach will be to take a network with a certain degree of periodicity, obtain an explicit handle for its \acl{PoA} over a compact interval, and then tessellate this behavior over the entire traffic spectrum $(0,\infty)$.
To carry this out, let $\game_{\totrate}$ be a nonatomic routing game consisting of a single \ac{OD} pair with traffic inflow $\totrate$.
This traffic is to be routed over the three-link parallel graph of \cref{fig:parallel} with cost functions
\begin{subequations}
\label{eq:cost-ex}
\begin{flalign}
\cost_{1}(\load_{1})
	&= \load_{1}^{\degr} \, \bracks*{1 + \tfrac{1}{2} \sin(\log\load_{1})},
	\\
\cost_{2}(\load_{2})
	&= \load_{2}^{\degr},
	\\
\cost_{3}(\load_{3})
	&= \load_{3}^{\degr} \, \bracks*{1 + \tfrac{1}{2} \cos(\log\load_{3})},
\end{flalign}
\end{subequations}
where $\degr$ is a positive integer.
It is easy to see that these cost functions are increasing, strictly convex and smooth on $[0,\infty)$ for all $\degr\geq 2$, and they all grow as $\Theta(\load^{\degr})$ at both traffic limits ($\load\to0$ and $\load\to\infty$).
Furthermore, the functions $\load_{\edge} \cost_{\edge}(\load_{\edge})$ are \emph{strictly} convex, so the optimum traffic allocation problem \eqref{eq:SO} admits a unique solution.
Hence, the only way for the game's \acl{PoA} to be equal to $1$ is when the game's (also unique) \acl{WE} coincides with the network's \acl{SO} flow.

As we show in \cref{app:example}, the equations determining the network's equilibrium and optimum flows never admit a common solution, so the \acl{PoA} is strictly greater than $1$ over any compact interval (showing in this way that the conclusion of \cref{prop:BPR-light,prop:PoA-equal-one} already fails for this example).
Moreover, the trigonometric terms in \eqref{eq:cost-ex} imply that these equations are periodic in a logarithmic scale (\ie in $\log\totrate$).
Hence, combining these two properties, we obtain:

\begin{proposition}
\label{prop:example}
In the three-link parallel network defined above,
$\PoA(\game_{\totrate})$ is periodic in $\log\totrate$ and oscillates between two values strictly greater than $1$.
\end{proposition}

\citet{ColComSca:SAGT2016} already provided examples of networks where $\limsup_{\totrate\to\infty}\PoA(\game_{\totrate}) >1$ but the cost functions involved were fairly irregular and the $\liminf$ of the \acl{PoA} was still $1$ (\ie selfish routing was still efficient infinitely often).
By contrast, in the above example, the \acl{PoA} is bounded away from $1$ for \emph{all} possible demands, and this despite the fact that the network's cost functions are smooth, strictly convex and grow as $\Theta(\load^{\degr})$ at both ends of the congestion spectrum.
This is a considerable sharpening of the example of \citet{ColComSca:SAGT2016} as it shows that there are cases where efficiency is \emph{never} achieved at equilibrium \textendash\ not even asymptotically.

\section{Networks with a single \ac{OD} pair}
\label{sec:single}

Despite the highly smooth and convex structure of the example network of \cref{prop:example}, closer inspection reveals that the growth rate of its cost functions exhibits persistent oscillations at both $0$ and $\infty$.
This naturally leads to the following question:
\emph{Does selfish routing remain bad for ``reasonable'' cost functions that do not behave irregularly in the limit?}

To quantify \textendash\ and discard \textendash\ such irregularities, we will employ the seminal notion of \emph{regular variation} (recalled below).
For clarity and concision, we will focus for now on networks with a single \ac{OD} pair;
the case of multiple \ac{OD} pairs will be discussed in detail later, in \cref{sec:multi,sec:variable}.

\subsection{Regular variation and edge classification}
\label{sec:regvar}

To present a unified perspective, we will tackle both ends of the congestion spectrum simultaneously by introducing the \emph{traffic limit indicator} $\pole \in \{0,\infty\}$:
letting $\totrate\to\pole$ gives the \emph{light traffic limit} for $\pole=0$ and the \emph{heavy traffic limit} for $\pole=\infty$.
Regular variation at either limit is then defined as follows:

\begin{definition}
\label{def:regvar}
A function $\test\from (0,\infty) \to (0,\infty)$ is said to be \emph{regularly varying at $\pole$} if
\begin{equation}
\label{eq:regvar}
\lim_{t\to\pole} \frac{\test(t\load)}{\test(t)}
	\quad
	\text{is finite and nonzero for all $\load>0$}.
\end{equation}
\end{definition}

In words, regular variation means that $\test(t)$ grows at the same rate when viewed at different scales (determined here by $\load$).
The concept itself dates back to \citet{Kar:MC1930,Kar:BSMF1933} and has been used extensively in functional analysis, probability, and large deviations theory \citep[see \eg][]{deHFer:Springer2006,JesMik:PIMB2006,Res:Springer2007};
for a comprehensive survey we refer the reader to \cite{BinGolTeu:CUP1989}.

Standard examples of regularly varying functions include all affine functions, polynomials, logarithms, and, more generally, all analytic functions (barring those with an essential singularity at $\pole=\infty$).%
\footnoteOR{Recall here that a function $\test(\load)$ is analytic on a domain $U$ if it is equal to its Taylor series on $U$.}
On the other hand, despite being bounded from above and below as $\Theta(\load^{\degr})$, the oscillatory cost functions \eqref{eq:cost-ex} used in the counterexample of \cref{sec:example} are \emph{not} regularly varying.
Indeed, at either $\pole=0$ or $\pole=\infty$, the limit
\begin{equation}
\lim_{t\to\pole} \frac{\cost_{1}(t\load)}{\cost_{1}(t)}
	= \lim_{t\to\pole} \frac{1 + \frac{1}{2} \sin(\log t + \log\load)}{1 + \frac{1}{2} \sin(\log t)} \load^{\degr}
\end{equation}
does not exist in $(0,\infty)$ unless $\log\load = k\pi$ for some $k\in\Z$ (and likewise for $\cost_{3}$).
In this way, regular variation provides a much finer view than polynomial growth.

With all this at hand, we will dispose of growth irregularities like the above by positing that the network's cost functions can be compared asymptotically to some regularly varying function $\cost(\load)$.
Specifically, given an ensemble of cost functions $\costs = \{\cost_{\edge}\}_{\edge\in\edges}$, we will say that a regularly varying function $\cost\from(0,\infty)\to(0,\infty)$ is a \emph{benchmark for $\costs$ at $\pole$} if the following \textpar{possibly infinite} limit exists for all $\edge\in\edges$
\begin{equation}
\label{eq:index}
\ind_{\edge}
	= \lim_{\load\to\pole} \frac{\cost_{\edge}(\load)}{\cost(\load)}.
\end{equation}
This limit will be called the \emph{index of edge $\edge$ at $\pole$}, and $\edge$ will be called \emph{fast}, \emph{slow}, or \emph{tight} (relative to $\cost$ at $\pole$) if $\ind_{\edge}$ is respectively $0$, $\infty$, or in-between.
In particular, when $\edge$ is tight, $\cost_{\edge}(\load)$  is also regularly varying and exhibits the same asymptotic behavior as the benchmark function $\cost(\load)$ at $\pole$;
if $\edge$ is fast, then $\cost_{\edge}(\load) = o(\cost(\load))$;
and, finally, if $\edge$ is slow, then $\cost(\load) = o(\cost_{\edge}(\load))$.
As such, a benchmark function groups the network's edges into three equivalence classes that exhibit the same qualitative behavior with respect to $\cost(\load)$.

Of course, this partition depends on the chosen benchmark and the traffic limit (light or heavy):
for instance, $\load^{2}$ is fast with respect to $\load$ at $0$, but it is slow at $\infty$.
For concision, we will not keep track of this dependence explicitly and instead rely on the context to resolve any ambiguities.
However, it will be important to keep in mind that the classification of fast and slow edges could be flipped when transitioning from heavy to light traffic and vice versa.

Now, since bottlenecks along a path are caused by its slowest edges, we also define the \emph{index of a path $\route\in\routes$} as
\begin{equation}
\label{eq:index-path}
\ind_{\route}
	= \max_{\edge\in\route} \ind_{\edge},
\end{equation}
and we say that $\route$ is \emph{fast}, \emph{slow}, or \emph{tight} based on whether $\ind_{\route}$ is $0$, $\infty$, or in-between.
Finally, given that traffic will tend to be routed along the fastest paths in the network, we define the \emph{index of the network} as
\begin{equation}
\label{eq:index-network-single}
\ind
	=\min_{\route\in\routes} \ind_{\route},
\end{equation}
and we say that the network is itself \emph{tight} if $0<\ind<\infty$.
In words, a path is fast (resp. tight/slow) if its slowest edge is fast (resp. tight/slow), and a network is tight if its fastest path is tight.

Defined this way, tightness guarantees that the network admits a path whose cost behaves asymptotically as a (positive) multiple of the benchmark function $\cost(\load)$.
The importance of this requirement is again illustrated by the cost model \eqref{eq:cost-ex} of the previous section:
if we only assumed that the network admits a path whose cost behaves as $\Theta(\cost(\load))$, then we would not be able to rule out the  pathological oscillations of the example in \cref{sec:example}.

\subsection{The light traffic limit}
\label{sec:single-light}

Thanks to the above legwork, we are in a position to state our main result for lightly congested networks with a single \ac{OD} pair:

\begin{theorem}
\label{thm:single-light}
Let $\game_{\totrate}$ be a nonatomic routing game with a single \ac{OD} pair.
If the network is tight under light traffic \textpar{$\pole=0$}, then
\begin{equation}
\lim_{\totrate\to0} \PoA(\game_{\totrate})
	= 1.
\end{equation}
\end{theorem}

In words, \cref{thm:single-light} simply states that if the cost of the network's fastest path is regularly varying at $0$, selfish routing becomes efficient in light traffic.
To streamline our presentation, \cref{thm:single-light} is proved in \cref{app:convergence} as a special case of a much more general statement.
Here, we focus on some immediate corollaries thereof:

\begin{corollary}
\label{cor:single-light-power}
Suppose that, for every edge $\edge\in\edges$, the limit $\lim_{\load\to0} \cost_{\edge}(\load) / \load^{\ord_{\edge}}$ is finite and nonzero for some $\ord_{\edge}\geq0$.
Then, $\PoA(\game_{\totrate}) \to 1$ as $\totrate\to0$.
\end{corollary}

\begin{Proof}
Referring to $\ord_{\edge}$ as the \emph{order} of $\edge$, define the order of a path $\route\in\routes$ as $\ord_{\route} = \min_{\edge\in\route} \ord_{\edge}$ and that of the network as $\ord = \max_{\route\in\routes} \ord_{\route}$.
Clearly,
$\lim_{\load\to0} \cost_{\edge}(\load)/\load^{\ord} = 0$ if and only if $\ord_{\edge}>\ord$;
$\lim_{\load\to0} \cost_{\edge}(\load)/\load^{\ord} = \infty$ if and only if $\ord_{\edge}<\ord$;
and
$\lim_{\load\to0} \cost_{\edge}(\load)/\load^{\ord} \in(0,\infty)$ if and only if $\ord_{\edge}=\ord$.
This shows that the network is tight with respect to $\cost(\load) = \load^{\ord}$ at $0$, so \cref{thm:single-light} applies.
\end{Proof}

\begin{corollary}
\label{cor:single-light-anal}
In a single \ac{OD}-network with analytic costs 
we have $\PoA(\game_{\totrate})\to1$ as $\totrate\to0$.
\end{corollary}

\begin{Proof}
If $\cost_{\edge}(\load) = \sum_{\pol=0}^{\infty} \cost_{\pol,\edge} \load^{\pol}$ for small enough $\load$, we have $\lim_{\load\to\infty} \cost_{\edge}(\load)/\load^{\ord_{\edge}} \in (0,\infty)$ for $\ord_{\edge} = \min\setdef{\pol\in\N}{\cost_{\pol,\edge}\neq0}$.
Our claim then follows from \cref{cor:single-light-power}.
\end{Proof}

\begin{corollary}
\label{cor:single-light-poly}
In a single \ac{OD}-network with polynomial costs
we have $\PoA(\game_{\totrate})\to1$ as $\totrate\to0$.
\end{corollary}

Of the above results, \cref{cor:single-light-anal,cor:single-light-poly} are of special practical interest because most latency models that have been proposed in the literature are polynomial or analytic at $0$.
In urban networks, the golden standard is the \ac{BPR} quartic model $\cost_{\edge}(\load) = a_{\edge} + b_{\edge} \load^{4}$, while basically all of the established queueing models used in the theory of packet-switched networks ($M/M/1$, $M/G/k$, $M/M/c$, etc.) are analytic at $0$ \citep{BG92}.

Despite appearances, the very wide applicability of \cref{thm:single-light} and its corollaries is fairly surprising.
Indeed, at first sight, one would expect that when $\totrate\to0$, traffic is so light that it doesn't really matter how it is routed.
This is indeed the case if, for instance, all paths in the network exhibit different positive costs when $\totrate=0$ (cf.~\cref{prop:PoA-equal-one}).
However, if the cost of an empty path is zero, this is no longer the case:
the optimum and equilibrium traffic assignments could be fairly different (even when the network is lightly congested), so there is no a priori reason for the price of anarchy to converge to $1$ as $\totrate\to0$ (the example of Section 3 clearly illustrates this phenomenon).
\cref{thm:single-light} shows that all that is needed for this to occur is for the network's cost functions to be faithfully represented by a common benchmark function:
when this condition is met, optimum and equilibrium costs no longer fluctuate but, instead, they converge to the same value.

\subsection{The heavy traffic limit}
\label{sec:single-heavy}

Our main result for highly congested networks with a single \ac{OD} pair is as follows: 

\begin{theorem}
\label{thm:single-heavy}
Let $\game_{\totrate}$ be a nonatomic routing game with a single \ac{OD} pair. 
If the network is tight under heavy traffic \textpar{$\pole=\infty$}, then
\begin{equation}
\lim_{\totrate\to\infty} \PoA(\game_{\totrate})
	= 1.
\end{equation}
\end{theorem}

In words, \cref{thm:single-heavy} simply states that if the cost of the network's fastest path is regularly varying at $\infty$, selfish routing becomes efficient in heavy traffic.
To compare and contrast the light and heavy traffic regimes, we relegate the proof of \cref{thm:single-heavy} to \cref{app:convergence} and only focus here on some immediate corollaries thereof:

\begin{corollary}
\label{cor:single-heavy-bounded}
Suppose there exists a path $\route\in\routes$ with bounded costs, that is, $\lim_{\load\to\infty}\cost_\edge(\load)<\infty$ for all $\edge\in\route$.
Then, $\PoA(\game_{\totrate}) \to 1$  as $\totrate\to\infty$.
\end{corollary}

\begin{Proof}
Taking $\cost(\load) = 1$, we get $\ind_{\edge} = \lim_{\load\to\infty} \cost_{\edge}(\load)\in(0,\infty]$ for all $\edge\in\edges$.
By assumption, there exists a path such that $0 < \ind_{\route} < \infty$, so \cref{thm:single-heavy} applies.
\end{Proof}

\begin{corollary}
\label{cor:single-heavy-power}
Suppose that the limit $\lim_{\load\to\infty} \cost_{\edge}(\load) / \load^{\ord_{\edge}}$ is finite and nonzero for some $\ord_{\edge}\geq0$ and all $\edge\in\edges$.
Then, $\PoA(\game_{\totrate}) \to 1$ as $\totrate\to\infty$.
\end{corollary}

\begin{Proof}
Shadowing the proof of \cref{cor:single-light-power}, let $\ord_{\route} = \max_{\edge\in\route} \ord_{\edge}$ and $\ord = \min_{\route\in\routes} \ord_{\route}$ (but note the reversal of the $\max$ and $\min$ operators).
Clearly,
$\lim_{\load\to\infty} \cost_{\edge}(\load)/\load^{\ord} = 0$ if and only if $\ord_{\edge}<\ord$;
$\lim_{\load\to\infty} \cost_{\edge}(\load)/\load^{\ord} = \infty$ if and only if $\ord_{\edge}>\ord$;
finally,
$\lim_{\load\to\infty} \cost_{\edge}(\load)/\load^{\ord} \in(0,\infty)$ if and only if $\ord_{\edge}=\ord$.
This shows that the network is tight with respect to $\cost(\load) = \load^{\ord}$ at $\infty$, so \cref{thm:single-heavy} applies.
\end{Proof}

\begin{corollary}
\label{cor:single-heavy-poly}
In a single \ac{OD}-network with polynomial costs
we have $\PoA(\game_{\totrate})\to1$ as $\totrate\to\infty$.
\end{corollary}

In a certain, precise sense, \cref{thm:single-light,thm:single-heavy} show that the high and low congestion regimes can be seen as different sides of the same coin.
By excluding pathological oscillations at either end of the congestion spectrum, regular variation ensures asymptotic regularity and guarantees that selfish routing becomes efficient in the limit:
specifically, tightness at $0$ guarantees efficiency in light traffic while tightness at $\infty$ guarantees efficiency in heavy traffic.
By this token, taking \cref{cor:single-light-poly,cor:single-heavy-poly} in tandem implies that selfish routing becomes efficient under both light and heavy traffic in networks with polynomial costs and a single \ac{OD} pair.

That being said, there are still important, quantitative differences between the light and heavy traffic limits.
For instance, even though \cref{cor:single-heavy-power,cor:single-heavy-poly} are direct analogues of their light traffic counterparts, the conclusion of \cref{cor:single-heavy-bounded} is false in light traffic (the three-link network of \cref{sec:example} serves again as a counterexample).
In fact, even in the case of polynomial costs (\cref{cor:single-light-poly} vs. \cref{cor:single-heavy-poly}), there is an important reversal of roles that takes place between fast and slow edges.
Specifically, edges that are fast in light traffic typically become slow under heavy traffic and vice versa (importantly however, tight edges are not re-classified under this regime change).
Nevertheless, despite this reversal, the \acl{PoA} still goes to $1$ in both cases.

\section{Networks with multiple \ac{OD} pairs}
\label{sec:multi}

We now extend our analysis to networks with multiple \ac{OD} pairs.
In this case, if the inflow rate of the $\pair$-th \ac{OD} pair is $\rate^{\pair}$, the total traffic inflow in the network is given by
\begin{equation}
\label{eq:rate-total}
\totrate
	= \sum_{\pair\in\pairs} \rate^{\pair},
\end{equation}
and we  write
\begin{equation}
\label{eq:relrate}
\relrate^{\pair}
	= \frac{\rate^{\pair}}{\totrate}
\end{equation}
for the \emph{relative inflow rate} of the $\pair$-th \ac{OD} pair \textendash\ \ie the fraction of the total traffic generated by the pair in question.
In the rest of this section, we will assume that the relative inflow of every \ac{OD} pair $\pair\in\pairs$ is a fixed positive 
constant that does not depend on $\totrate$;
the case of variable inflow rates will be discussed in detail in \cref{sec:variable}.

The key difference with the single-pair setting is that routing costs for different \ac{OD} pairs may exhibit completely different asymptotic behaviors in the limit.
As a result, in the presence of multiple \ac{OD} pairs, the definition of the network's index (and the related notion of tightness) must be re-examined.
To do so, given that the traffic generated by an \ac{OD} pair will tend to be routed along the pair's fastest path, we first define the \emph{index of an \ac{OD} pair} $\pair\in\pairs$ as
\begin{equation}
\label{eq:index-pair}
\ind^{\pair}
	= \min_{\route\in\routes^{\pair}} \ind_{\route}.
\end{equation}
Just like edges and paths, this index can be used to classify \ac{OD} pairs as \emph{fast}, \emph{slow} or \emph{tight} depending on whether $\ind^{\pair}$ is respectively $0$, $\infty$, or in-between.
The \emph{index of the network} is then defined as
\begin{equation}
\label{eq:index-network-multi}
\ind
	= \max_{\pair\in\pairs} \ind^{\pair},
\end{equation}
and we say that the network is \emph{tight} if $0<\ind<\infty$.
Heuristically, this definition simply captures the fact that the leading contribution to congestion is due to the ``costliest'' \ac{OD} pairs in the network;
obviously, if there is but a single \ac{OD} pair, this last point is moot and \eqref{eq:index-network-multi} reduces to \eqref{eq:index-network-single}.


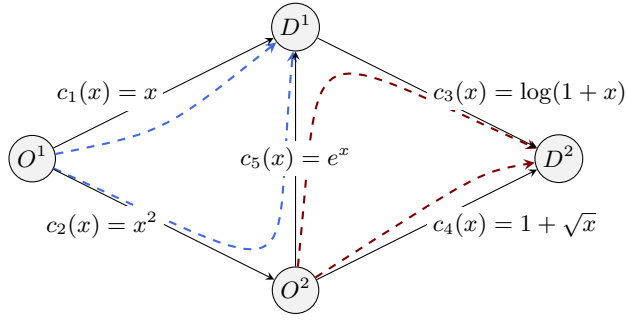
\begin{figure}
\centering
\footnotesize

\begin{tikzpicture}
[scale=3.5,
nodestyle/.style={circle,draw=black,fill=gray!10, inner sep=1pt},
edgestyle/.style={-},
>=stealth]

\small

\coordinate (A) at (-1,0);
\coordinate (B) at (1,0);
\coordinate (C) at (0,1/2);
\coordinate (D) at (0,-1/2);

\node (A) at (A) [nodestyle] {$\source^{1}$\vphantom{bp}};
\node (C) at (C) [nodestyle] {$\sink^{1}$\vphantom{bp}};
\node (D) at (D) [nodestyle] {$\source^{2}$\vphantom{bp}};
\node (B) at (B) [nodestyle] {$\sink^{2}$\vphantom{bp}};

\draw[dashed,thick,RoyalBlue,->] (A) .. controls ($0.5*(A)+0.7*(C)+(0,-0.25)$) .. 
node[near start, below] {} (C);
\draw[dashed,thick,RoyalBlue,->] (A) .. controls ($(D)+0.1*(B)$) and ($(D)+0.1*(A)$)..
node[near start, above] {} (C);

\draw[dashed,thick,Maroon,->] (D) .. controls ($0.6*(D)+0.7*(B)+(0,0.25)$) .. 
node[near start, below] {} (B);
\draw[dashed,thick,Maroon,->] (D) .. controls ($(C)-0.1*(A)$) and ($(C)+0.1*(D)$)..
node[near start, above] {} (B);

\draw [edgestyle,->] (A) to node [midway, left, fill=white, inner sep=2pt] {$\cost_{1}(\load) = \load$} (C);
\draw [edgestyle,->] (A) to node [midway, left, fill=white, inner sep=2pt] {$\cost_{2}(\load) = \load^{2}$} (D);
\draw [edgestyle,->] (C) to node [midway, right, fill=white, inner sep=2pt] {$\cost_{3}(\load) = \log(1+\load)$} (B);
\draw [edgestyle,->] (D) to node [midway, right, fill=white, inner sep=2pt] {$\cost_{4}(\load) = 1+\sqrt{\load}$} (B);
\draw [edgestyle,->] (D) to node [midway, fill=white, inner sep=2pt] {$\cost_{5}(\load) = e^{\load}$} (C);

\end{tikzpicture}
\caption{A Wheatstone network with two \ac{OD} pairs (cf.~\cref{ex:benchmarks} below).
In heavy traffic, the network is tight relative to the benchmark function $\cost(\load)=\load$;
in light traffic, the network is tight relative to the benchmark $\cost(\load) = 1$.}
\label{fig:Wheatstone}
\vspace{-1em}
\end{figure}


\begin{example}
\label{ex:benchmarks}
To illustrate the above concepts, consider a Wheatstone network with two \ac{OD} pairs and cost functions as in \cref{fig:Wheatstone}.
Focusing first on the heavy traffic limit, the benchmark $\cost(\load) = \load$ would classify edge $1$ as tight, edges $2$ and $5$ as slow, and edges $3$ and $4$ as fast.
Accordingly, the first \ac{OD} pair would be classified as tight while the second \ac{OD} pair would be classified as fast;
since no pair is slow and at least one pair is not fast, the network is itself tight.

In the light traffic limit, the same benchmark would classify edges $1$ and $3$ as tight, edges $4$ and $5$ as slow, and edge $2$ as fast.
Under this classification, the first \ac{OD} pair would again be tight, but the second \ac{OD} pair would now be classified as slow (because all its paths contain a slow edge), so the network would no longer be tight.
A moment's reflection shows that the reason for this is that the benchmark function $\cost(\load) = \load$ is not well-suited for the second \ac{OD} pair.
Instead, if we take the benchmark $\cost(\load) = 1$, the first \ac{OD} pair would be classified as fast (because it has a fast path, namely edge $1$) and the second pair would be classified as tight, so the network would now be tight.
For a systematized version of this benchmark selection procedure, see the proof of \cref{cor:multi-ratios} below.
\end{example}

With all this at hand, our next result states that if the  costliest \ac{OD} pair in the network admits a tight path, selfish routing becomes asymptotically efficient in the limit:

\begin{theorem}
\label{thm:multi}
Let $\game_{\totrate}$ be a nonatomic routing game.
If the network is tight in the limit as $\totrate\to\pole$, then
\begin{equation}
\lim_{\totrate\to\pole} \PoA(\game_{\totrate})
	= 1.
\end{equation}
\end{theorem}

In words, if
\begin{inparaenum}
[(\itshape a\upshape)]
\item
every \ac{OD} pair has a path which is not slow,
and
\item
the fastest path of the slowest \ac{OD} pair has a regularly varying cost,
\end{inparaenum}
selfish routing becomes efficient in the limit.
Motivated by the strong connection between \cref{thm:single-light,thm:single-heavy}, \cref{thm:multi} has been stated in a way that does not discriminate between the light and heavy traffic regimes.
The reason for this is to highlight the role of the tightness assumption:
tightness at $0$ guarantees efficiency in light traffic ($\pole=0$) while tightness at $\infty$ guarantees efficiency in heavy traffic ($\pole=\infty$).

Of course, both \cref{thm:single-light,thm:single-heavy} follow as corollaries of \cref{thm:multi} by taking respectively $\pole=0$ or $\infty$ and specializing to a single \ac{OD} pair (in which case \cref{eq:index-network-single,eq:index-network-multi} coincide).
Other than that, however, the same caveats apply regarding the passage from light to heavy traffic:
the classification of fast and slow edges could be reversed, the equilibrium/\acl{SO} flows could be drastically different in the two regimes, etc.
To illustrate all this, we proceed with some further corollaries of \cref{thm:multi} (which we prove in \cref{app:convergence}):

%

\begin{corollary}
\label{cor:multi-ratios}
If the network's costs are regularly varying at $\pole$ and the \textpar{possibly infinite} limit
\begin{equation}
\label{eq:edgeratio}
\ind_{\edge,\edgealt}
	= \lim_{\load\to\pole} \cost_{\edge}(\load)/\cost_{\edgealt}(\load)
\end{equation}
exists for all $\edge,\edgealt\in\edges$, then $\PoA(\game_{\totrate}) \to 1$ as $\totrate\to\pole$.
\end{corollary}

\begin{Proof}
Define a total preorder among the network's edges by setting $\edge\mleq\edgealt$ if and only if $\ind_{\edge,\edgealt} \leq 1$.
For each path $\route\in\routes$, choose a maximal element $\edge_{\route}$ of $\route$,
\ie an edge $\edge_{\route}\in\route$ such that $\edge\mleq \edge_{\route}$ for all $\edge\in\route$.
Then, for each \ac{OD} pair $\pair\in\pairs$, choose a path $\route^{\pair}$ for which $\edge_{\route^{\pair}}$ is minimal, \ie $\edge_{\route^{\pair}} \mleq \edge_{\routealt}$ for all $\routealt\in\routes^{\pair}$.
Finally, pick an \ac{OD} pair $\pair\in\pairs$ such that $\edge_{\route^{\pair}}$ is maximal, \ie $\edge_{\route^{\pairalt}} \mleq \edge_{\route^{\pair}}$ for all $\pairalt\in\pairs$.
Setting $\eq\edge = \edge_{\route^{\pair}}$ and taking $\cost(\load) \equiv \cost_{\eq\edge}(\load)$ as a benchmark, it is easy to verify that the network is tight at $\pole$, so \cref{thm:multi} applies.
\end{Proof}

\medskip
The proof of \cref{cor:multi-ratios} shows that the ``comparison index'' $\ind_{\edge,\edgealt}$ induces a preference relation which refines the coarser classification of the network's edges into fast, slow and tight.
Of course, this ordering could be reversed when passing from light to heavy traffic, but the existence thereof (along with regular variation) guarantees that the \acl{PoA} is asymptotically equal to $1$ in both cases.

In addition to the above, \cref{cor:multi-ratios} also gives an alternative way to prove the following analogue of \cref{cor:single-light-power,cor:single-heavy-power}:

\begin{corollary}
\label{cor:multi-power}
Suppose that the limit $\lim_{\load\to\pole} \cost_{\edge}(\load) / \load^{\ord_{\edge}}$ is finite and nonzero for some $\ord_{\edge}\geq0$ and all $\edge\in\edges$.
Then, $\PoA(\game_{\totrate}) \to 1$ as $\totrate\to\pole$.
\end{corollary}

\begin{Proof}
Observe that
\begin{equation}
\frac{\cost_{\edge}(\load)}{\cost_{\edgealt}(\load)}
	= \frac{\cost_{\edge}(\load)}{\load^{\ord_{\edge}}}
	\frac{\load^{\ord_{\edge}}}{\load^{\ord_{\edgealt}}}
	\frac{\load^{\ord_{\edgealt}}}{\cost_{\edgealt}(\load)},
\end{equation}
so $\lim_{\load\to\pole} \cost_{\edge}(\load)/\cost_{\edgealt}(\load)$ exists for all $\edge\in\edges$.
Our claim then follows from \cref{cor:multi-ratios}.
\end{Proof}

\smallskip
We thus obtain the following important corollary for polynomial cost functions:

\begin{corollary}
\label{cor:multi-poly}
In networks with polynomial costs, $\PoA(\game_{\totrate})\to1$ as $\totrate\to\pole$.
\end{corollary}


\medskip
In words, \cref{cor:multi-poly} yields the general principle that we stated in the introduction:

\vspace{.5ex}
\begin{center}
\itshape
In networks with polynomial costs,\\
the \acl{PoA} becomes $1$ under both light and heavy traffic.
\end{center}
\vspace{.5ex}
We find this principle particularly appealing as it indicates that the \acl{PoA} may attain high values \emph{only} in an intermediate regime (where the traffic is neither light nor heavy).

\section{Networks with variable inflow rates}
\label{sec:variable}

An important assumption in the analysis of the previous section is that the relative inflow rate $\relrate^{\pair} = \rate^{\pair}/\totrate$ of each \ac{OD} pair $\pair\in\pairs$ does not fluctuate in the limit \textendash\ \ie all pairs are assumed to generate a constant fraction of the overall traffic.
In general however, this assumption need not hold:
for instance, in an urban road network, central \ac{OD} pairs generate disproportionately more traffic during rush hour than peripheral, suburban destinations, so it is not reasonable to assume that traffic scales up maintaining a constant traffic ratio between different \ac{OD} pairs.

To understand the impact of this variability, consider two independent links, $\edge_{1}$ and $\edge_{2}$, with corresponding cost functions $\cost_{1}(\load) = \load$ and $\cost_{2}(\load) = \load^{2}$.
Suppose further that these links are joining two uncoupled \ac{OD} pairs with inflow rates $\rate^{1}$ and $\rate^{2}$ and total inflow $\totrate = \rate^{1}+\rate^{2}$.
If both inflows scale in the limit as $\Theta(\totrate)$, the cost of the first pair will scale as $\Theta(\totrate)$ while that of the second pair will scale as $\Theta(\totrate^{2})$.
As such, the leading contribution to congestion will be that of the first \ac{OD} pair in light traffic, and that of the second pair in heavy traffic.
If, however, the inflow of the first pair scales as $\Theta(\totrate)$ but that of the second pair scales as $\Theta(\totrate^{1/3})$, the induced costs will scale respectively as $\Theta(\totrate)$ and $\Theta(\totrate^{2/3})$;
consequently, the most costly \ac{OD} pair will now be the second one in light traffic and the first one in heavy traffic.

This reversal of roles shows that the asymptotic behavior of the relative inflow rates $\relrate^{\pair} = \rate^{\pair}/\totrate$ could end up painting a completely different picture in the limit.
In particular, if these relative rates oscillate wildly in the limit, the \acl{PoA} may exhibit a likewise irregular asymptotic behavior, even if the underlying network is well-behaved (for instance, even if it is tight; cf.~\cref{ex:inefficient} below).
As a result, special care must be taken to define and study the asymptotic regime in networks with variable traffic demands.

To that end, let $\game_{\run}$ be a sequence of nonatomic routing games with total inflow $\totrate_{\run} = \sum_{\pair\in\pairs} \rate_{\run}^{\pair}$ induced by a sequence of inflow rates $\rate_{\run}^{\pair}$ for each \ac{OD} pair $\pair\in\pairs$.
The light and heavy traffic limits are obviously recovered depending on whether the total inflow $\totrate_{\run}$ converges respectively to $\pole=0$ or $\pole=\infty$ as $\run\to\infty$.
However, the relative inflow rates $\relrate_{\run}^{\pair} = \rate_{\run}^{\pair}/\totrate_{\run}$ could now exhibit very different behaviors as $\run\to\infty$:
in particular, as discussed above, the relative inflow of an \ac{OD} pair could oscillate \textendash\ or even vanish \textendash\ in the limit.
To capture such phenomena, we introduce below the notion of \emph{salience}:

\begin{definition}
\label{def:salient}
Let $\game_{\run}$ be a sequence of nonatomic routing games with relative inflow rates $\relrate_{\run}^{\pair}$, $\pair\in\pairs$.
We  say that a subset $\pairsalt\subseteq\pairs$ of \ac{OD} pairs is \emph{salient} if
\begin{equation}
\label{eq:salient}
\liminf_{\run\to\infty} \sum_{\pair\in\pairsalt} \relrate_{\run}^{\pair}
	> 0,
\end{equation}
\ie if the total fraction of the traffic generated by the \ac{OD} pairs in $\pairsalt$ is non-negligible in the limit.
\end{definition}

Obviously, if the sequence of relative inflow vectors $\relrate_{\run} = (\relrate_{\run}^{\pair})_{\pair\in\pairs}$ converges to a well-defined limit, $\pairsalt$ will be salient if and only if some \ac{OD} pair of $\pairsalt$ is itself salient \textendash\ \ie if and only if $\liminf_{\run\to\infty} \relrate_{\run}^{\pair} > 0$ for some $\pair\in\pairsalt$.
However, if this is not the case, a set of \ac{OD} pairs may be salient even if none of its constituent pairs is salient:
for instance, if there are two \ac{OD} pairs, ``$+$'' and ``$-$'', with relative inflows $\relrate_{\run}^{\pm} = (1 \pm (-1)^{\run})/2$, neither pair is salient but their union \emph{is} (since $\relrate_{\run}^{+} + \relrate_{\run}^{-} = 1$ for all $\run$).
Thus, the notion of salience does not rule out fluctuations in the relative inflows of \emph{individual} \ac{OD} pairs;
it only posits that the set of \ac{OD} pairs in question carries enough traffic in the limit.

Bearing all this in mind, our main result for networks with variable inflow rates is as follows:

\begin{theorem}
\label{thm:multi-variable}
Let $\game_{\run}$ be a sequence of nonatomic routing games with inflow rates $\rate_{\run}^{\pair}$ and total inflow $\totrate_{\run} = \sum_{\pair\in\pairs} \rate_{\run}^{\pair}$.
Suppose further that:
\begin{enumerate}
[\indent \upshape(\itshape a\upshape)]
\item
\label{cond:limit}
Traffic is either light or heavy in the limit, \ie $\lim_{\run\to\infty} \totrate_{\run} = \pole\in\{0,\infty\}$.
\item
\label{cond:nonslow}
Every \ac{OD} pair has a path which is not slow, \ie $\ind^{\pair}<\infty$ for all $\pair\in\pairs$.
\item
\label{cond:salient}
The set of tight \ac{OD} pairs is salient, \ie $\liminf_{\run\to\infty} \sum_{\pair:\ind^{\pair}>0} \relrate_{\run}^{\pair} > 0$.
\end{enumerate}
Then, $\PoA(\game_{\run}) \to 1$ as $\run\to\infty$.
\end{theorem}

Heuristically, \condref{cond:limit} above simply indicates the traffic regime under study (light or heavy), whereas \condref{cond:nonslow} guarantees that the network's benchmark function correctly classifies the paths that are not too costly in each \ac{OD} pair.
Finally, \condref{cond:salient} guarantees that tight \ac{OD} pairs are indeed relevant in terms of traffic, \ie they account for a non-negligible fraction of the total inflow.

In view of all this, \cref{thm:multi-variable} can be seen as a direct extension of our ``fixed-rate'' analysis in \cref{sec:single,sec:multi}:
indeed, in the case of constant (positive) relative inflows, salience boils down to asking that the network admits at least one tight \ac{OD} pair, so \cref{thm:multi} can be obtained as a special case of \cref{thm:multi-variable}.
Below, we provide some further corollaries of \cref{thm:multi-variable} along these lines:

\begin{corollary}
If every \ac{OD} pair in the network is tight, then $\PoA(\game_{\run})\to1$.
\end{corollary}

\begin{corollary}
If the network is tight and every \ac{OD} pair is salient, then $\PoA(\game_{\run})\to 1$.
\end{corollary}

On the other hand, it is worth noting that if salience fails, we can draw no definitive conclusions for the \acl{PoA}.
We illustrate the main reasons for this via two examples below:


\begin{figure}[t]
\centering
\footnotesize

\begin{tikzpicture}
[scale=1.1,
nodestyle/.style={circle,draw=black,fill=gray!10, inner sep=1pt, text = MyBlue},
edgestyle/.style={-},
>=stealth]

\coordinate (A) at (-\textwidth/7,0);
\coordinate (B) at (\textwidth/7,0);
\coordinate (C) at (-\textwidth/7,-1.5);
\coordinate (D) at (\textwidth/7,-1.5);

\node (A) at (A) [nodestyle] {$\source^{1}\vphantom{bp}$};
\node (B) at (B) [nodestyle] {$\sink^{1}\vphantom{bp}$};
\node (C) at (C) [nodestyle] {$\source^{2}\vphantom{bp}$};
\node (D) at (D) [nodestyle] {$\sink^{2}\vphantom{bp}$};
\node (phantom) at (0,-2) {};

\draw [edgestyle,->] (A) to [bend left = 25] node [midway, fill=white] {$\cost_{1}(\load) = 1$} (B);
\draw [edgestyle,->] (A) to [bend right = 25] node [midway, fill=white] {$\cost_{2}(\load) = \load$} (B);
\draw [edgestyle,->] (C) to node [midway,fill=white] {$\cost_{3}(\load) = 0$} (D);

\end{tikzpicture}
\caption{A simple network with two uncoupled \ac{OD} pairs.}
\label{fig:uncoupled}
\vspace{-1em}
\end{figure}
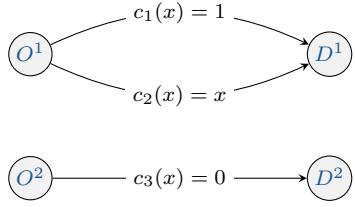


\begin{example}[Efficiency without salience]
\label{ex:efficient}
Consider the simple network of \cref{fig:uncoupled}, where two ``uncoupled'' \ac{OD} pairs respectively encounter a standard Pigou network and an independent link with zero cost.
In heavy traffic, the benchmark function $\cost(\load) = 1$ classifies the \ac{OD} pair $\pair=1$ as tight and the pair $\pair=2$ as fast, so the network is itself tight.
Note also that the second \ac{OD} pair does not affect the network's \acl{PoA} because it has a single routing option and its cost is identically equal to $0$;
however, it still affects the definition of relative inflows.

If we take the inflow sequence $\rate_{\run}^{1} = \sqrt{\run}$ and $\rate_{\run}^{2} = \run$, we get $\totrate_{\run} = \run+\sqrt{\run}\to\infty$ and $\relrate_{\run}^{1}\to 0$ as $\run\to\infty$, so the first \ac{OD} pair is not salient.
Since the second \ac{OD} pair is not tight, \condref{cond:salient} fails;
nevertheless, if we apply \cref{thm:single-heavy} to the first \ac{OD} pair by itself, we obtain $\lim_{\totrate\to\infty} \PoA(\game_{\totrate}) = 1$ (recall here that the second pair does not affect the network's \ac{PoA}).
In other words, \condref{cond:salient} is not necessary for selfish routing to be asymptotically efficient.
\end{example}

\begin{example}[Inefficiency without salience]
\label{ex:inefficient}
Consider the same network as above but take the periodically oscillating inflow sequence
\begin{equation}
\rate_{\run}^{1}
	=
	\begin{cases}
	1
		&\text{for $\run$ odd},
	\\
	1+2\run
		& \text{for $\run$ even},
\end{cases}
\qquad
\text{and}
\qquad
\rate_{\run}^{2}
	= \begin{cases}
	2\run
		&\text{for $\run$ odd},
		\\
	0
		&\text{for $\run$ even}.
\end{cases}
\end{equation}
We then have $\totrate_{\run} = 2\run+1\to\infty$ and $\liminf_{\run\to\infty} \relrate_{\run}^{1} = 0$ so, again, \condref{cond:salient} fails (but in a different way).
This time, whenever $\run$ is odd, the network's \acl{PoA} is equal to that of a Pigou network with inflow $1$ (because the second \ac{OD} pair is costless).
Thus, for $\run$ odd, we get $\PoA(\game_{\run}) = 4/3$, which is the worst-case value for networks with affine costs;
as such, the conclusion of \cref{thm:multi-variable} does not hold in general if we just drop the salience condition.
\end{example}

The above examples suggest that there is a finer mechanism at work which is not captured by the intersection of tightness and salience.
At a high level, the crucial component of this mechanism seems to be that asymptotic efficiency is guaranteed if the network remains tight after suitably modifying the network's cost functions to take into account the scaling of the inflow of each \ac{OD} pair.
However, getting an exact statement at this level of generality is fairly cumbersome, so we do not attempt it here.

\section{Rate of convergence}
\label{sec:rate}

The analysis of the previous sections provides a wide range of sufficient conditions guaranteeing that selfish routing becomes efficient in the limit;
however, it does not provide any indication for the rate at which the network's \acl{PoA} converges to $1$.
In this section, we derive such rates (including subleading terms) for networks with polynomial costs of the form
\begin{equation}
\label{eq:polynomials}
\cost_{\edge}(\load)
	= \sum_{\pol=\ord_{\edge}}^{\degr_{\edge}}\cost_{\edge,\pol}\load^{\pol},
\end{equation}
where
all coefficients are assumed nonnegative ($\cost_{\edge,\pol}\geq0$)
and
$\ord_{\edge}$ and $\degr_{\edge}$ respectively denote the smallest and largest powers present
(so $\cost_{\edge,\ord_{\edge}}, \cost_{\edge,\degr_{\edge}}>0$);
by convention, we also take $\ord_{\edge}=\infty$ and $\degr_{\edge}=0$ when $\cost_{\edge}(\load)\equiv 0$.%
\footnoteOR{This follows the standard \textendash\ if somewhat surprising at first \textendash\ convention that $\sup\varnothing =-\infty$, $\inf\varnothing=\infty$.}
This model covers in particular the \ac{BPR} ``constant plus monomial'' model \eqref{eq:BPR} but also extends more easily to networks with more intricate cost functions.

In contrast to our qualitative analysis, the two traffic limits (light and heavy) exhibit different quantitative behaviors, so we treat them separately.

\subsection{The light traffic case}
\label{sec:rate-light}

We begin with the light traffic limit ($\pole=0$).
To motivate our analysis, we start with a simple example of a Pigou network with monomial costs as shown in \cref{fig:Pigou-poly}:
for $\degr_{1},\degr_{2}>0$, the zero-flow travel time of both links is zero, so \cref{prop:PoA-equal-one} does not apply.
Instead, as we show below, the network's \acl{PoA} decays to $1$ following a power law:


\begin{figure}[t]
\centering
\footnotesize

\begin{tikzpicture}
[scale=1.1,
nodestyle/.style={circle,draw=black,fill=gray!10, inner sep=1pt, text = MyBlue},
edgestyle/.style={-},
>=stealth]

\coordinate (A) at (-\textwidth/7,0);
\coordinate (B) at (\textwidth/7,0);

\node (A) at (A) [nodestyle] {$\source\vphantom{bp}$};
\node (B) at (B) [nodestyle] {$\sink\vphantom{bp}$};
\node (phantom) at (0,-2) {};

\draw [edgestyle,->] (A) to [bend left = 25] node [midway, fill=white] {$\cost_{1}(\load) = \load^{\degr_{1}}$} (B);
\draw [edgestyle,->] (A) to [bend right = 25] node [midway, fill=white] {$\cost_{2}(\load) = \load^{\degr_{2}}$} (B);

\end{tikzpicture}
\vspace{-4ex}
\caption{A two-link Pigou network with monomial costs.}
\label{fig:Pigou-poly}
\vspace{-2ex}
\end{figure}
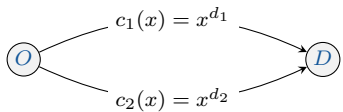


\begin{proposition}
\label{prop:Pigou-light}
Consider a two-link parallel network with cost functions $\cost_{1}(\load) = \load^{\degr_{1}}$ and $\cost_{2}(\load) = \load^{\degr_{2}}$, $0<\degr_{1} < \degr_{2}$, and a single \ac{OD} pair with inflow $\totrate$.
Then
\begin{equation}
\label{eq:Pigou-asym-light}
\PoA(\game_{\totrate})=1+\pigourate\totrate^{\adegr}+o(\totrate^{\adegr})
\end{equation}
where
\begin{flalign}
\label{eq:Pigou-exp-light}
\adegr
	&=\degr_{2}/\degr_{1} - 1,
\intertext{and}
\label{eq:Pigou-light}
\pigourate
	&= \degr_{1} \parens*{\frac{1+\degr_{2}}{1+\degr_{1}}}^{1+1/\degr_{1}} - \degr_{2}
	> 0.
\end{flalign}
\end{proposition}


\begin{figure*}[t]
\centering
\footnotesize
\subfigure{\includegraphics[width=.475\textwidth]{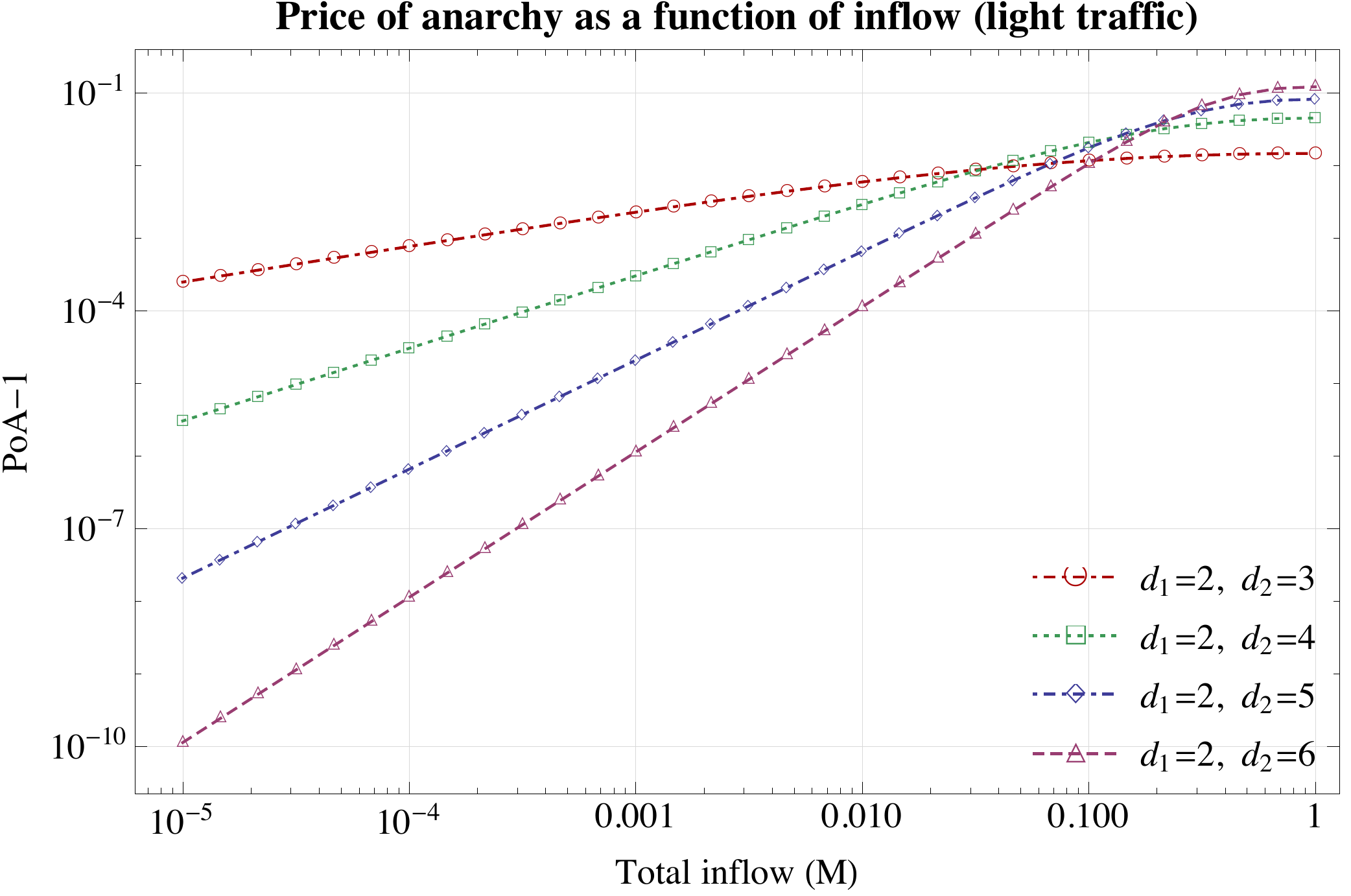}}
\hfill
\subfigure{\includegraphics[width=.475\textwidth]{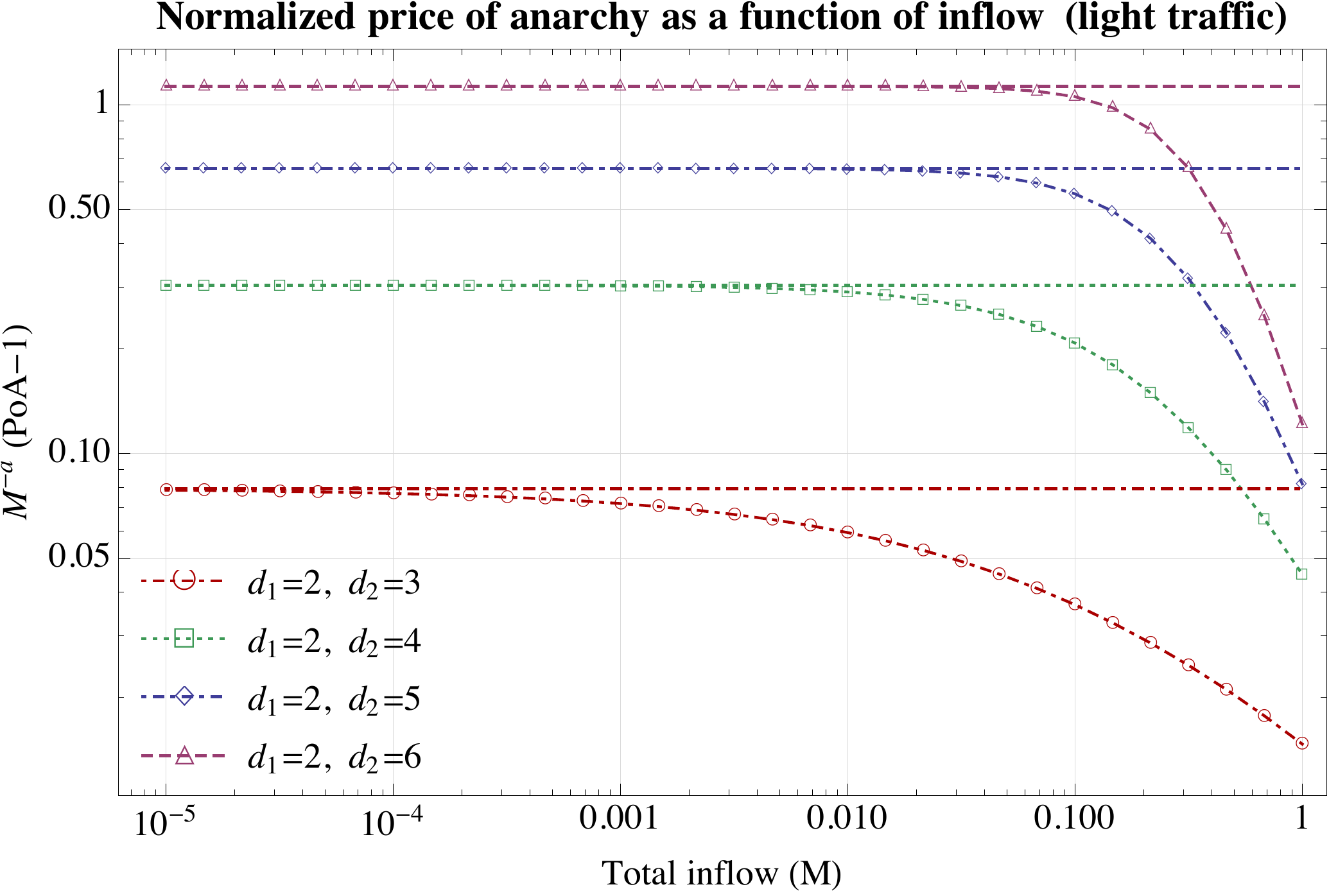}}
\caption{The rate of convergence of the \acl{PoA} in a lightly congested Pigou network as in \cref{fig:Pigou-poly}.
The figure on the left shows the \acl{PoA} as a function of the total traffic inflow $\totrate$ for different values of the 
exponents $\degr_1$ and $\degr_2$.
In the figure to the right, the \acl{PoA} has been rescaled by $\totrate^{-\adegr}$ with $a=\degr_2/\degr_1-1$, showing that $\PoA(\game_{\totrate}) \sim 1 + \pigourate\totrate^{\adegr}$ for some $\pigourate>0$;
the horizontal asymptotes correspond precisely to the expression \eqref{eq:Pigou-light} for $\pigourate$.}
\label{fig:Pigou-light}
\end{figure*}


In words, \cref{prop:Pigou-light} shows that the rate of convergence of the \acl{PoA} is controlled by the ratio $\degr_{2}/\degr_{1}$:
the largest the ratio of degrees, the fastest the decay of the \acl{PoA} (for a numerical illustration, see \cref{fig:Pigou-light}).
This behavior is consistent with \cref{prop:PoA-equal-one} which predicts that $\PoA(\game_{\totrate}) \equiv 1$ if $\totrate$ is small enough and $\degr_{1} = 0$.
Indeed, taking $\degr_{1}\to0$ in \eqref{eq:Pigou-asym-light} shows that $\PoA(\game_{\totrate}) = 1 + \bigoh(\totrate^{\adegr})$ for any $\adegr>0$, suggesting in turn that the rate of decay of $\PoA(\game_{\totrate})$ is qualitatively different in this case.

Another case worth noting is when $\degr_{1} = \degr_{2}$, \ie when both links are equivalent in terms of performance.
In this case, $\PoA(\game_{\totrate})$ is identically equal to $1$ for all values of $\totrate$ (\cref{prop:BPR-light} already guarantees as much when $\totrate$ is not large).
However, \eqref{eq:Pigou-asym-light} would seem to suggest that the \acl{PoA} can remain large as $\totrate\to0$ (since $\adegr = 1 - 1 = 0$ when $\degr_{1} = \degr_{2}$).
The solution of this apparent paradox is provided by looking at the multiplicative constant $\pigourate$:
when $\degr_{1} = \degr_{2}$, we also have $\pigourate=0$, so the resulting contribution to the \acl{PoA} is $0$ \textendash\ not $\Omega(1)$.

The above highlights the importance of the relative rate of decay of the network's edge costs as a function of the inflow.
Since monomials with lower exponents are more costly in the low traffic limit, this rate is dominated by the smallest power in \eqref{eq:polynomials}.
Thus, motivated by the index machinery of \cref{sec:single,sec:multi}, we respectively define the \emph{order} of an edge $\edge\in\edges$, that of a path $\route\in\routes$, of an \ac{OD} pair $\pair\in\pairs$, and that of the network itself as
\begin{subequations}
\label{eq:ord-light}
\begin{flalign}
\label{eq:ord-edge-light}
\ord_{\edge}
	&= \min_{\vphantom{\pol>0}}\setdef{\pol}{\cost_{\edge,\pol} > 0}
	\\
\label{eq:ord-path-light}
\ord_{\route}
	&= \min_{\edge\in\route}\ord_{\edge},
	\\
\label{eq:ord-pair-light}
\ord^{\pair}
	&= \max_{\route\in\routes^{\pair}} \ord_{\route},
	\\
\label{eq:ord-net-light}
\ord
	&= \min_{\pair\in\pairs} \ord^{\pair}.
\end{flalign}
\end{subequations}
In view of the above, the network is tight with respect to the benchmark function $\cost(\load)=\load^{\ord}$, and an edge $\edge\in\edges$ is fast when $\ord_\edge>\ord$, tight when $\ord_\edge=\ord$, and slow if $\ord_\edge<\ord$.
We then denote the set of the network's slow edges as
\begin{flalign}
\label{eq:slow-light}
\slow\edges
	&= \setdef{\edge\in\edges}{\ord_{\edge} < \ord},
\intertext{and we write}
\label{eq:gap-light}
\slow\ord
	&= \max_{\edge\in\slow\edges} \ord_{\edge}
\end{flalign}
for the order of the fastest edge in $\slow\edges$ (again employing the standard convention that $\max\varnothing = -\infty$, so $\slow\ord = -\infty$ when there are no slow edges).

Building on the intuition gained from \cref{prop:Pigou-light}, our main quantitative result for low traffic is that the network's \ac{PoA} decays to $1$ following a power law that depends only on the ratio between the order of the network ($\ord$) and that of its fastest slow edge ($\slow\ord$):

\begin{theorem}
\label{thm:rate-light}
Let $\game_{\totrate}$ be a nonatomic routing game with polynomial costs, total inflow $\totrate$, and fixed relative inflows.
Then, there exist non-negative constants $\firstconst\geq 0$ and $\secondconst\geq 0$ such that
\begin{equation}
\label{eq:rate-light}
\PoA(\game_{\totrate})
	\leq 1 + \firstconst\totrate + \secondconst\totrate^{\adegr},
\end{equation}
where $\adegr = \ord/\slow\ord - 1$ and $\secondconst=0$ whenever $\slow\edges=\varnothing$.
\end{theorem}

\cref{thm:rate-light} was stated for networks with fixed relative inflows for simplicity only:
in \cref{app:rate}, we state and prove a more general result for networks with variable relative inflows as in \cref{sec:multi}.
In terms of intuition, we only note here that \cref{thm:rate-light} complements the insights gained from \cref{prop:BPR-light,prop:PoA-equal-one} in an important way:
when there is no single ``best path'' under zero inflow, the network's \acl{PoA} is no longer identically equal to $1$ for small inflows but instead behaves as a power law.

\subsection{The heavy traffic case}
\label{sec:rates-heavy}

We now turn to the heavy traffic limit ($\pole=\infty$).
As in the light traffic case, we start with a simple \textendash\ but illuminating \textendash\ example of a two-link Pigou network where precise results can be obtained:


\begin{figure*}[t]
\centering
\footnotesize
\subfigure{\includegraphics[width=.475\textwidth]{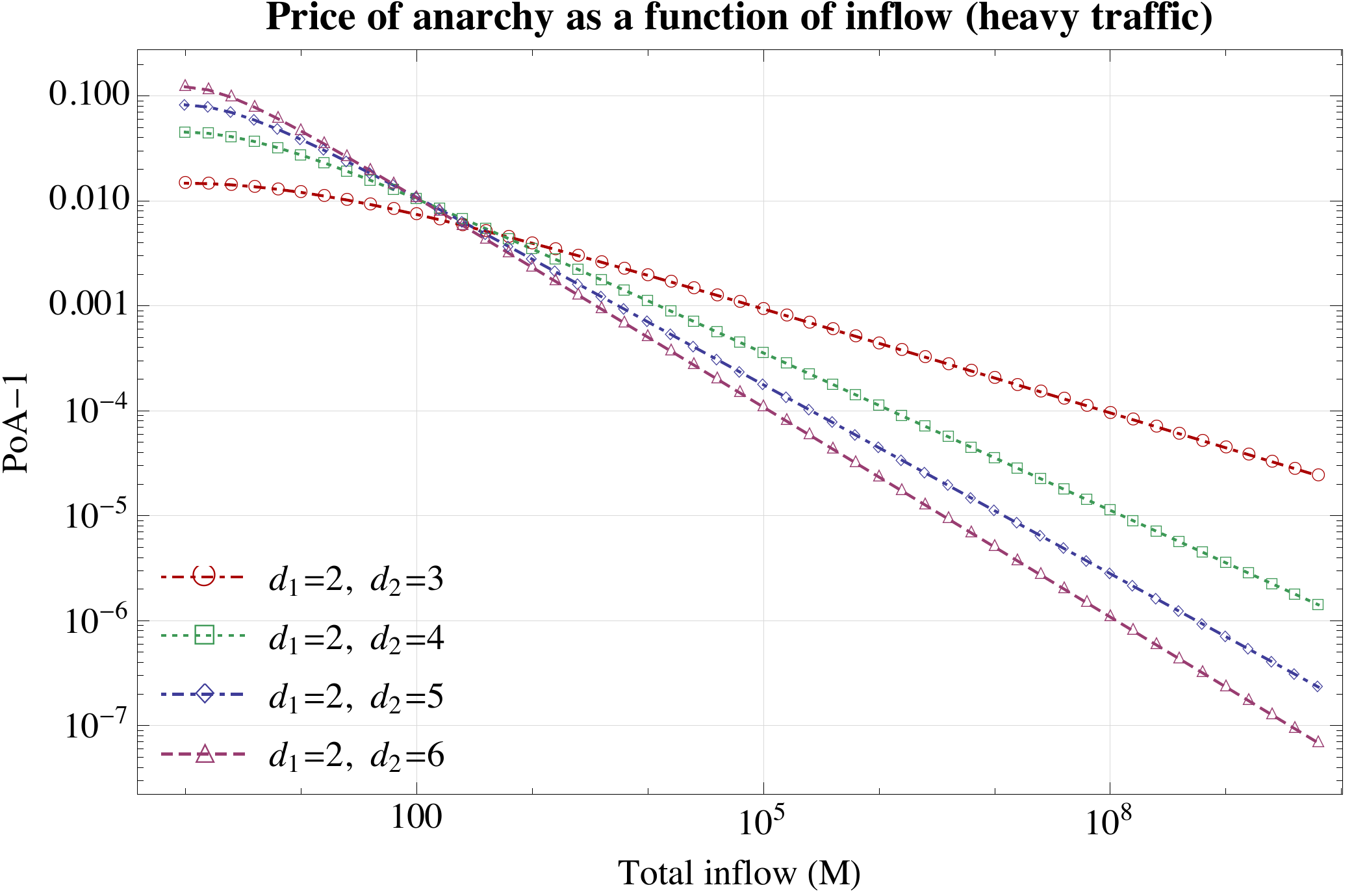}}
\hfill
\subfigure{\includegraphics[width=.475\textwidth]{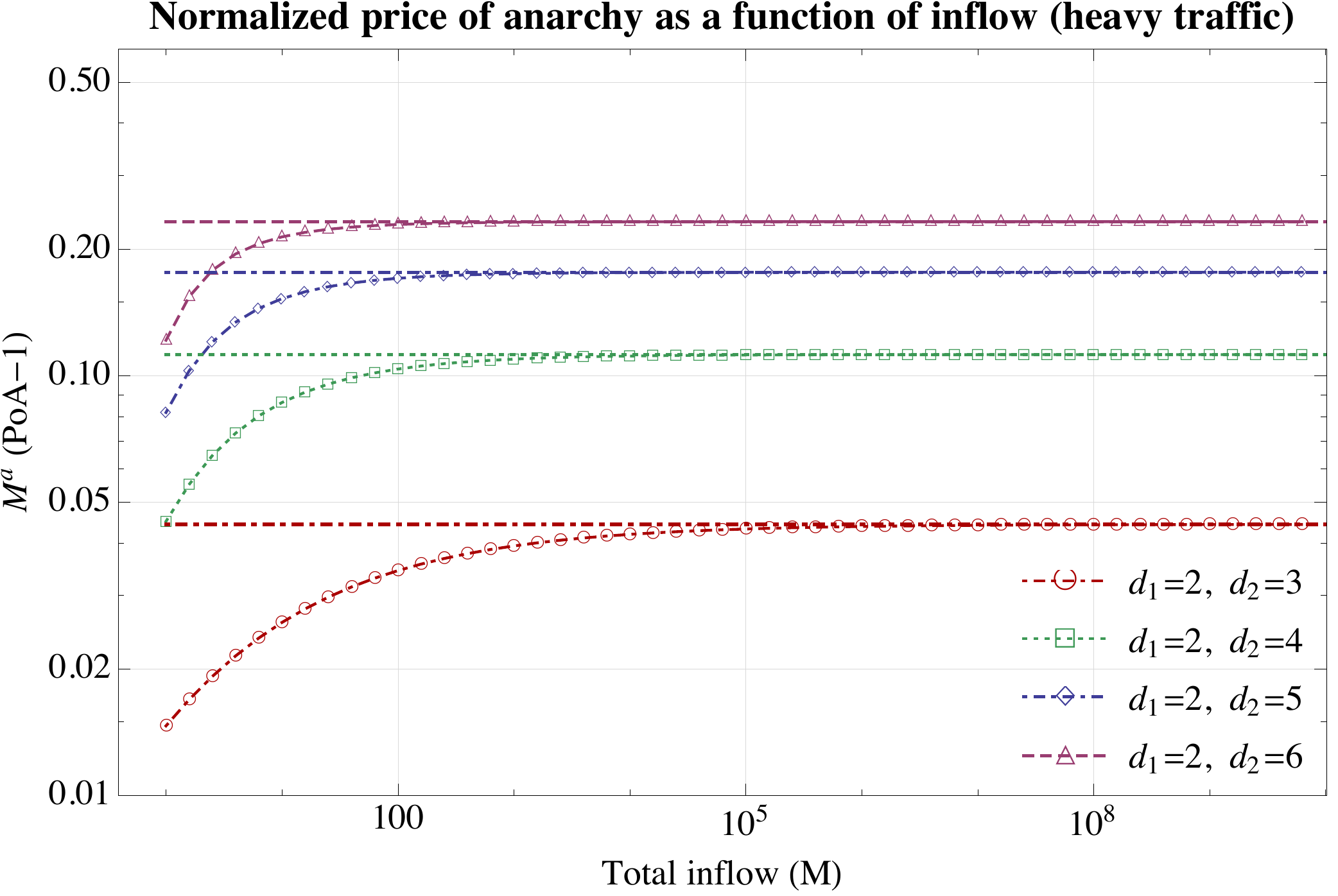}}
\caption{The rate of convergence of the \acl{PoA} in a heavily congested Pigou network as in \cref{fig:Pigou-poly}.
The figure on the left shows the \acl{PoA} as a function of the total traffic inflow $\totrate$ for different values of the exponents $\degr_1$ and $\degr_2$.
In the figure to the right, the \acl{PoA} has been rescaled by $\totrate^{\adegr}$ with $a=1-\degr_1/\degr_2$, showing that $\PoA(\game_{\totrate}) \sim 1 + \pigourate/\totrate^{\adegr}$ for some $\pigourate>0$;
the horizontal asymptotes correspond precisely to the expression \eqref{eq:Pigou-heavy} for $\pigourate$.
}
\vspace{-1ex}

\label{fig:Pigou-heavy}
\end{figure*}


\begin{proposition}
\label{prop:Pigou-heavy}
Consider a two-link parallel network with cost functions $\cost_{1}(\load) = \load^{\degr_{1}}$ and $\cost_{2}(\load) = \load^{\degr_{2}}$, $0<\degr_{1} < \degr_{2}$, and a single \ac{OD} pair with inflow $\totrate$.
Then
\begin{equation}
\label{eq:Pigou-asym-heavy}
\PoA(\game_{\totrate})=1+\pigourate\totrate^{-\adegr}+o(\totrate^{-\adegr})
\end{equation}
where
\begin{flalign}
\label{eq:Pigou-exp-heavy}
\adegr
	&= 1 - \degr_{1}/\degr_{2}
\intertext{and}
\label{eq:Pigou-heavy}
\pigourate
	&= \degr_{2} \parens*{\frac{1 + \degr_{1}}{1 + \degr_{2}}}^{1+1/\degr_{2}} - \degr_{1}
	> 0.
\end{flalign}
\end{proposition}

For illustration purposes, we plotted in \cref{fig:Pigou-heavy} the asymptotic behavior of the network's \acl{PoA} for different values of $\degr_{1}$ and $\degr_{2}$.
In the same vein as in the light traffic limit, two special cases that are of interest here are when $\degr_{2}=\infty$ and when $\degr_{1} = \degr_{2}$.
In the former ($\degr_{2}=\infty$), \cref{eq:Pigou-exp-heavy} gives $\adegr=1$, indicating a convergence rate of the order of $\bigoh(1/\totrate)$:
since $\adegr<1$ for all finite $\degr_{2}$, this is the best possible rate that can be achieved in the heavy traffic limit.
For the latter ($\degr_{1} = \degr_{2}$), \cref{eq:Pigou-exp-heavy} gives $\adegr=0$, suggesting that the \acl{PoA} can remain large as $\totrate\to\infty$.
This seems to be inconsistent with the fact that the network's \acl{PoA} is identically equal to $1$ when $\degr_{1} = \degr_{2}$ but a closer look reveals that the multiplicative constant $\pigourate$ of \eqref{eq:Pigou-heavy} is also $0$ when $\degr_{1} = \degr_{2}$, thus reconciling the two results.

Now, even though the above result does not apply to more general networks with polynomial costs, it still highlights the main mechanism at play.
Specifically, for large edge loads $\load$, the dominant term in \eqref{eq:polynomials} is the one with highest degree $\degr_\edge$. As we've discussed before, this indicates a complete reversal of roles between light and heavy traffic:
for $\degr_{1} < \degr_{2}$,  $\load^{\degr_{1}}$ is \emph{slower} than $\load^{\degr_{2}}$ when $\load\to 0$, but \emph{faster}  when $\load\to\infty$. 
Thus, with an obvious adaptation of what we did for light traffic, we define the \emph{order} of an edge $\edge\in\edges$, that of a path $\route\in\routes$, of an \ac{OD} pair $\pair\in\pairs$, and of the network itself as
\begin{subequations}
\label{eq:ord-heavy}
\begin{flalign}
\label{eq:ord-edge-heavy}
\degr_{\edge}
	&= \max_{\vphantom{\pol>0}} \setdef{\pol}{\cost_{\edge,\pol} > 0}
	\\
\label{eq:ord-path-heavy}
\degr_{\route}
	&= \max_{\edge\in\route}\degr_{\edge},
	\\
\label{eq:ord-pair-heavy}
\degr^{\pair}
	&= \min_{\route\in\routes^{\pair}} \degr_{\route},
	\\
\label{eq:ord-net-heavy}
\degr
	&= \max_{\pair\in\pairs} \degr^{\pair}.
\end{flalign}
\end{subequations}
With all this at hand, we see that the network is tight with respect to the benchmark $\cost(\load)=\load^{\degr}$,
so an edge $\edge\in\edges$ is fast when $\degr_{\edge}<\degr$, tight when $\degr_{\edge} = \degr$, and slow if $\degr_{\edge} > \degr$.
The set of the network's slow edges is then denoted as
\begin{flalign}
\label{eq:slow-heavy}
\slow\edges
	&= \setdef{\edge\in\edges}{\degr_{\edge} > \degr},
\intertext{and we write}
\label{eq:gap-heavy}
\slow\degr
	&= \min_{\edge\in\slow\edges} \degr_{\edge}
\end{flalign}
for the order of the fastest edge in $\slow\edges$ (again employing the standard convention that $\min\varnothing = \infty$, so $\slow\degr = \infty$ when there are no slow edges).

Mutatis mutandis, this definition is the same as in light traffic except for a reversal of the max/min operators.
Our main result for heavily congested networks confirms this intuition:

\begin{theorem}
\label{thm:rate-heavy}
Let $\game_{\totrate}$ be a nonatomic routing game with polynomial costs, total inflow $\totrate$, and fixed relative inflows.
Then, there exist non-negative constants $\firstconst\geq 0$ and $\secondconst\geq 0$ such that
\begin{equation}
\label{eq:rate-heavy}
\PoA(\game_{\totrate})
	\leq 1 + \frac{\firstconst}{\totrate} + \frac{\secondconst}{\totrate^{\adegr}},
\end{equation}
where $\adegr=1-\degr/\slow\degr$ and  $\secondconst=0$ whenever $\slow\edges=\varnothing$.
\end{theorem}


\begin{remark}
\label{rem:samedeg}
As in the light traffic case, if the costs are monomials of the same degree, then $\firstconst=\secondconst=0$ and $\PoA(\game_{\totrate})\equiv 1$ for all $\totrate>0$.
\end{remark}

In words, given that $\adegr < 1$ when there is at least one slow edge in the network ($\adegr=1$ and $\secondconst = 0$ otherwise), \cref{thm:rate-heavy} states the network's \acl{PoA} converges to $1$ as $\bigoh(1/\totrate^{\adegr})$ with an $\bigoh(1/\totrate)$ subleading term.
In particular, in the presence of a single slow edge $\edge$ with $\degr_{\edge} > \degr$, the convergence exponent $\adegr$ in \eqref{eq:rate-heavy} can become as small as $1/(\degr+1)$, thus pointing to a slower convergence rate in networks with  routing costs of high degree and a small gap between the degree of tight and slow edges.
On the other hand, if there are no slow edges  we have $\slow\edges = \varnothing$ and  we get an $\bigoh(1/\totrate)$ rate of convergence.

On this issue, \citet{WuMohChe:ArXiv2017} recently showed that if all the network's cost functions are of the \ac{BPR} type $\cost_\edge(\load)=a_\edge+b_\edge\load^\degr$ with the 
same degree $\degr$, 
then $\PoA(\game_{\totrate}) = 1+\bigoh(\totrate^{-\degr})$ as $\totrate\to\infty$.
In this special case, the rate of convergence is faster than the prediction of \cref{thm:rate-heavy}, a gap which points to a sharp discontinuity that occurs when all costs have the same degree.
To see this in a concrete example, consider again the two-link Pigou network of \cref{fig:Pigou-poly}.
By symmetry, if $\degr_1=\degr_2$, the fraction of traffic routed on edge $1$ at optimum and at equilibrium coincide
\begin{equation}
\opt{\unitflow}_{1}
	= \eq{\unitflow}_{1}
	= \frac{1}{2}
	\quad
	\text{for all $\totrate > 0$},
\end{equation}
implying in turn that $\PoA(\game_{\totrate})$ is identically equal to $1$.
On the other hand, when $\degr_1< \degr_2$, both fractions $\opt{\unitflow}_{1}$ and $\eq{\unitflow}_{1}$ converge to $1$ as $\totrate\to\infty$.
\cref{prop:Pigou-heavy} shows that the rate of convergence of the \acl{PoA} in this case is exactly of order $\Theta(1/\totrate^{\adegr})$ and cannot be improved.

Put differently, \cref{prop:Pigou-heavy} shows that the slightest difference in edge degrees causes the rate of convergence of the \acl{PoA} to drop abruptly to $\Theta(\totrate^{-\adegr})$;
in fact, \cref{eq:Pigou-heavy} even provides an explicit expression for the proportionality constant in the high congestion rate $\Theta(\totrate^{-\adegr})$.
By this token, the bounds provided by \cref{thm:rate-heavy} are tight and cannot be improved in general, even in the class of two-link parallel networks with monomial costs.

\section{Discussion}
\label{sec:discussion}

Most of the literature on the \acl{PoA} \textendash\ for congestion games and not only \textendash\ has traditionally focused on worst-case upper bounds for different classes of networks, cost functions, and/or types of players.
Several of these results have become milestones in the field and have had a significant impact in practical considerations for traffic networks.
However, real-world situations involve a fixed network and traffic flows that are not necessarily close to these worst-case scenarios.
Thus, in addition to determining \emph{how} bad can selfish routing become, it is also important to determine \emph{when} these cases are relevant.

Our goal in this paper was to provide an answer to this question by examining the behavior of the \acl{PoA} at each end of the congestion spectrum.
Under fairly mild assumptions (that always include networks with polynomial costs), we found that the \ac{PoA} goes to $1$ in both cases, independently of the network's topology, and even when there are multiple \ac{OD} pairs.
What we find appealing about this result is that it is essentially independent of the underlying graph and/or the distribution of \ac{OD} pairs in the network.
Especially in the heavy traffic limit, this means that selfish routing is not the real cause of increased delays:
from a social planner's point of view, sophisticated tolling/rerouting schemes that target the optimum traffic assignment will not yield considerable gains over a ``laissez-faire'' approach where each traffic element takes the fastest available path.

%

\appendix

\section{Proofs of the results of \cref{sec:example}}
\label{app:example}

We start this appendix with the proofs of \cref{prop:PoA-equal-one,prop:BPR-light}.
To that end, we first establish the following result:
\begin{lemma}
\label{lem:argmin-set}
For sufficiently small $\totrate$, equilibrium and optimum traffic allocations only employ paths in $\argmincost^{\pair}=\argmin_{\route\in\routes^{\pair}}\cost_{\route}(0)$.
\end{lemma}

\begin{Proof}
In a slight abuse of notation, let $\cost_{\route}(\totrate)=\sum_{\edge\in\route}\cost_\edge(\totrate)$ denote the cost of the path $\route$ if all its edges carry load equal to the total inflow $\totrate$.
Clearly, if $\totrate$ is small enough, we have $\cost_{\routealt}(0) > \cost_{\route}(\totrate)$ for all $\route\in\argmincost^{\pair}$ and all $\routealt\in\routes^{\pair}\setminus\argmincost^{\pair}$.
Hence, for an equilibrium flow $\eq\flow$, we have
\begin{equation}
\cost_{\routealt}(\eq\flow)
	\geq \cost_{\routealt}(0)
	> \cost_{\route}(\totrate)
	\geq \cost_{\route}(\eq\flow),
\end{equation}
implying in turn that $\eq\flow_{\routealt}=0$.
Likewise, since an optimal flow $\opt\flow$ is an equilibrium for the marginal costs $\opt\cost_\edge(\load)=\cost_\edge(\load)+\load\;\cost'_\edge(\load)$ and $\opt\cost_{\routealt}(0) > \opt\cost_{\route}(\totrate)$, similar considerations show that an optimum flow profile cannot route any traffic along a path $\routealt\in\routes^{\pair}\setminus\argmincost^{\pair}$.
\end{Proof}

To proceed, it is more convenient to start with \cref{prop:PoA-equal-one}:

\begin{Proof}[Proof of \cref{prop:PoA-equal-one}]
By \cref{lem:argmin-set}, when $\totrate$ is small enough, both the equilibrium and the optimum must route the total inflow $\rate^\pair$ along the unique path in $\argmincost^{\pair}$. Hence the equilibrium and optimal flows coincide and therefore the \acl{PoA} is equal to $1$.
\end{Proof}

With this result at hand, we have:

\begin{Proof}[Proof of \cref{prop:BPR-light}]
If $\argmincost^{\pair}$ is a singleton for all $\pair\in\pairs$, our claim follows from \cref{prop:PoA-equal-one}. 
Otherwise, by \cref{lem:argmin-set}, if $\totrate$ is small enough, for every $\pair\in\pairs$, only paths in $\argmincost^{\pair}$ are used in equilibrium.
Moreover, if $\route,\routealt\in\argmincost^{\pair}$, then, for $\totrate$ small enough, we have
\begin{equation}
\sum_{\edge\in\route} a_{\edge} + b_{\edge} (\eq{\load}_{\edge})^{\degr}
	= \sum_{\edge\in\routealt} a_{\edge} + b_{\edge} (\eq{\load}_{\edge})^{\degr},
\end{equation}
and hence
\begin{equation}
\label{eq:BPReq}
\sum_{\edge\in\route} b_{\edge} (\eq{\load}_{\edge})^{\degr}
	= \sum_{\edge\in\routealt} b_{\edge} (\eq{\load}_{\edge})^{\degr}.
\end{equation}
Again, by \cref{lem:argmin-set}, if $\totrate$ is small enough, for every $\pair\in\pairs$, only paths in $\argmincost^{\pair}$ are used at the optimum. 
If $\route,\routealt\in\argmincost^{\pair}$, then, for $\totrate$ small enough, we have
\begin{equation}\label{}
\sum_{\edge\in\route} a_{\edge} + (\degr+1) b_{\edge} \opt{\load}_{\edge}^{\degr} = 
\sum_{\edge\in\routealt} a_{\edge} + (\degr+1) b_{\edge} \opt{\load}_{\edge}^{\degr},
\end{equation}
that is,
\begin{equation}\label{eq:BPRopt}
\sum_{\edge\in\route} (\degr+1) b_{\edge} \opt{\load}_{\edge}^{\degr} = 
\sum_{\edge\in\routealt} = (\degr+1) b_{\edge} \opt{\load}_{\edge}^{\degr}.
\end{equation}
Comparing \eqref{eq:BPReq} and \eqref{eq:BPRopt}, we see that the two equations are satisfied by the same loads. Therefore, for $\totrate$ small enough, there exist an equilibrium and an optimum having the same flows. Uniqueness of the equilibrium and optimum costs provides the result. 
\end{Proof}

We now present the proof of the counterexample with an oscillating \ac{PoA} of \cref{sec:counterexample}:

\begin{Proof}[Proof of \cref{prop:example}]
Since an unused edge has a cost of zero under \eqref{eq:cost-ex}, all three edges must be used at equilibrium.
Hence, for a given value of the total inflow $\totrate = \load_{1} + \load_{2} + \load_{3}$, the load profile $\load = (\load_{1},\load_{2},\load_{3})$ is a \acl{WE} if and only if $\cost_{1}(\load_{1}) = \cost_{2}(\load_{2}) = \cost_{3}(\load_{3})$.
In that case, the normalized profile $\unitload = \load/\totrate$ satisfies
\begin{flalign}
\label{eq:WE-ex}
\unitload_{1}^{\degr} \bracks*{1 + \tfrac{1}{2} \sin(\log \totrate\unitload_{1})}
	&= \unitload_{2}^{\degr}
	= \unitload_{3}^{\degr} \, \bracks*{1 + \tfrac{1}{2} \cos(\log \totrate\unitload_{3})}.
\intertext{Likewise, after differentiating and rearranging, the conditions for the network's \acl{SO} flow are}
\label{eq:SO-ex}
\unitload_{1}^{\degr} \bracks*{1 + \tfrac{1}{2} \sin(\log \totrate\unitload_{1}) + \tfrac{1}{2(\degr+1)} \cos(\log \totrate\unitload_{1})}
	&= \unitload_{2}^{\degr}
	= \unitload_{3}^{\degr} \, \bracks*{1 + \tfrac{1}{2} \cos(\log \totrate\unitload_{3}) - \tfrac{1}{2(\degr+1)} \sin(\log \totrate\unitload_{3})}.
\end{flalign}
We now show that \cref{eq:WE-ex,eq:SO-ex} never admit a common solution.
Indeed, this can occur if and only if
\begin{equation}
\cos(\log \totrate\unitload_{1})
	= 0
	= \sin(\log \totrate\unitload_{3}),
\end{equation}
\ie if and only if there exist integers $k_{1},k_{3}\in\Z$ such that
\begin{equation}
\begin{aligned}
\label{eq:k}
\log \totrate \unitload_{1}
	&= k_{1}\pi + \frac{\pi}{2},
	\\
\log \totrate\unitload_{3} 
	&= k_{3}\pi.
\end{aligned}
\end{equation}
This implies that $\sin(\log \totrate\unitload_{1}) = \pm1$ and $\cos(\log \totrate\unitload_{3}) = \pm1$, leading to the following cases:

\medskip
\paragraph{Case 1: $\sin(\log \totrate\unitload_{1}) = 1$, $\cos(\log \totrate\unitload_{3}) = 1$\afterhead}
Substituting in \eqref{eq:WE-ex} we get $\unitload_{1}^\degr = \unitload_{3}^\degr$ so \eqref{eq:k} gives
\begin{equation}
k_{1}\pi + \frac{\pi}{2}
	= k_{3}\pi.
\end{equation}
This gives $k_{3} - k_{1} = 1/2$, which cannot hold for integer values of $k_{1}$ and $k_{3}$.

\medskip
\paragraph{\itshape Case 2: $\sin(\log \totrate\unitload_{1}) = 1$, $\cos(\log \totrate\unitload_{3}) = -1$\afterhead}
As above, from \eqref{eq:WE-ex} we get $3\unitload_{1}^\degr = \unitload_{3}^\degr$, so \eqref{eq:k} gives
\begin{equation}\label{}
\frac{1}{d}\log 3 + k_{1}\pi + \frac{\pi}{2}
	= k_{3}\pi.
\end{equation}
This yields $\frac{\log 3}{\pi}=\degr(k_3-k_1-\frac{1}{2})$, which again cannot hold for $k_{1},k_{3},\degr \in\Z$.

\medskip
The remaining two cases lead to a contradiction in the same way, implying that the game's \acl{WE} and \acl{SO} flows cannot coincide for any value of $\totrate$.
Since \cref{eq:WE-ex,eq:SO-ex} are periodic in $\log\totrate$, it follows that the game's \acl{PoA} oscillates periodically at a logarithmic scale.
Thus, focusing on the period $1\leq\totrate\leq e^{2\pi}$, we conclude that
\begin{equation}
\inf_{\totrate>0} \PoA(\game_{\totrate})
	= \min_{1 \leq \totrate \leq e^{2\pi}} \PoA(\game_{\totrate})
	> 1,
\end{equation}
\ie the \aclp{WE} of the network in \cref{fig:parallel} remain strictly inefficient under both light and heavy traffic.
\end{Proof}

\section{Convergence of the \acl{PoA}}
\label{app:convergence}

We now prove \cref{thm:multi-variable};
\cref{thm:single-light,thm:single-heavy,thm:multi} will then follow as special cases of this more general result.
To that end, we begin with two auxiliary lemmas concerning the asymptotic behavior of regularly varying functions:

\begin{lemma}[\citealp{Kar:BSMF1933}]
\label{lem:Karamata}
If $\test$ is regularly varying at $\pole$, there exists some $\tdegr\in\R$ such that
\begin{equation}
\label{eq:reg-poly}
\lim_{t\to\pole} \frac{\test(tx)}{\test(t)}
	= x^{\tdegr}
	\quad
	\text{for all $x>0$}.
\end{equation}
\end{lemma}

\cref{lem:Karamata} is a classical result in the theory of regularly varying functions and gives rise to the term ``$\tdegr$-regularly varying'' for functions satisfying \eqref{eq:reg-poly};
for a proof, see, \eg \cite{BinGolTeu:CUP1989}.

The second lemma is a more technical asymptotic comparison result allowing us to replace a $\tdegr$-regularly varying function by a monomial of degree $\tdegr$ in the limit:

\begin{lemma}
\label{lem:reg-var-frac}
Let $\pole\in\{0,\infty\}$ and consider two functions $f,\test\from (0,\infty)\to (0,\infty)$ such that:
\begin{enumerate}
[\indent\upshape(1)]
\item
$f$ is nondecreasing.
\item
$\test$ is $\tdegr$-regularly varying at $\pole$ for some $\tdegr>0$.
\item
$\lim_{x\to\pole} f(x)/\test(x) = \ind\in[0,\infty)$.
\end{enumerate}
If $\totrate_{\run}\to\pole$ and $\unitload_{\run}\to \unitload\in [0,\infty)$, then
\begin{equation}
\lim_{\run\to\infty} \frac{f(\totrate_{\run}\unitload_{\run})}{\test(\totrate_{\run})}
	= \ind \unitload^{\tdegr}.
\end{equation}
\end{lemma}

\begin{Proof}
We first consider the case $\pole=\infty$.
If $\unitload>0$, the sequence $\load_{\run} = \totrate_{\run} \unitload_{\run}$ diverges to infinity, so our claim follows from Theorem~1.5.2 in \citet{BinGolTeu:CUP1989} by writing
\begin{equation}\label{eq:A2}
\frac{f(\totrate_{\run}\unitload_{\run})}{\test(\totrate_{\run})}
	= \frac{f(\load_{\run})}{\test(\load_{\run})}
	\cdot \frac{\test(\totrate_{\run}\unitload_{\run})}{\test(\totrate_{\run})}
	\to\ind \unitload^{\tdegr}.
\end{equation}
If $\unitload=0$, then, for all $\eps>0$, we have $\unitload_{\run}\leq\eps$ if $\run$ is sufficiently large.
Then, using the monotonicity of $f$ and the previous argument, we get
\begin{equation}
0
	\leq \limsup_{\run\to\infty} \frac{f(\totrate_{\run}\unitload_{\run})}{\test(\totrate_{\run})}
	\leq\limsup_{\run\to\infty} \frac{f(\totrate_{\run}\eps)}{\test(\totrate_{\run})}
	=\ind\eps^{\tdegr}.
\end{equation}
Taking $\eps\to 0$, we conclude that $f(\totrate_{\run}\unitload_{\run})/\test(\totrate_{\run})\to 0 = \ind \unitload^\tdegr$, as claimed.

The case $\pole=0$ is even simpler. Indeed, we now have that $\load_{\run} = \totrate_{\run} \unitload_{\run}$ tends to 0, so that the result follows using \eqref{eq:A2} and invoking Theorem~1.5.2 in \citet{BinGolTeu:CUP1989}.
\end{Proof}

Now, to proceed with the proof of \cref{thm:multi-variable}, we will require some additional notation.
First, fix some inflow vector $\rate = (\rate^{\pair})_{\pair\in\pairs}$ with total inflow $\totrate = \sum_{\pair\in\pairs} \rate^{\pair}$ and relative inflows
$\relrate = (\relrate^{\pair})_{\pair\in\pairs}$. Instead of working directly with the flow variables $\flow\in\flows$,
it will be more convenient to introduce the \emph{normalized traffic allocation} variables $\unitflow^{\pair} = (\unitflow_{\route}^{\pair})_{\route\in\routes^{\pair}}$ defined as
\begin{equation}
\label{eq:unitflow}
\unitflow_{\route}^{\pair}
	= \flow_{\route}/\rate^{\pair}
	\quad
	\text{for all $\route\in\routes^{\pair}$, $\pair\in\pairs$}.
\end{equation}
We clearly have $\sum_{\route\in\routes^{\pair}} \unitflow_{\route}^{\pair} = 1$ for all $\pair\in\pairs$;
we will also write $\unitflows^{\pair} = \simplex(\routes^{\pair})$ for the simplex of traffic allocations of $\pair\in\pairs$ and $\unitflows = \times_{\pair\in\pairs} \unitflows^{\pair}$ for the product thereof.
Moreover, descending to the edge level, we define the \emph{normalized load} induced by the $\pair$-th \ac{OD} pair on $\edge\in\edges$ as
\begin{equation}
\label{eq:unitload}
\unitload_{\edge}^{\pair}(\unitflow)
	= \sum_{\route\in\routes^{\pair}, \route\ni\edge} \unitflow_{\route}^{\pair}
\end{equation}
and we denote respectively the \emph{normalized} and \emph{total load} on edge $\edge\in\edges$ as
\begin{equation}
\label{eq:unitfunct}
\unitfunct_{\edge}(\unitflow,\relrate)
	= \text{$\sum_{\pair\in\pairs}$} \; \relrate^{\pair} \unitload_{\edge}^{\pair}(\unitflow)
	\quad
	\text{and}
	\quad
\load_{\edge}(\unitflow,\rate)
	= \totrate\;\unitfunct_{\edge}(\unitflow,\relrate)=\text{$\sum_{\pair\in\pairs}$} \;\rate^{\pair} \unitload_{\edge}^{\pair}(\unitflow).
\end{equation}
Finally, based on the index framework of \cref{sec:single,sec:multi}, we will respectively denote the set of the network's fast, tight and slow edges as
\begin{subequations}
\begin{alignat}{2}
&\fast\edges
	&\:=\:
	&\setdef{\edge\in\edges}{\ind_{\edge}=0},
	\\[2pt]
&\tight\edges
	&\:=\:
	&\setdef{\edge\in\edges}{0<\ind_{\edge}<\infty},
	\\[2pt]
&\slow\edges
	&\:=\:
	&\setdef{\edge\in\edges}{\ind_{\edge} = \infty},
\end{alignat}
\end{subequations}
and, in obvious notation, we will write \eg $\slow\routes$ for the set of the network's slow paths, $\tight\pairs$ for the set of tight \ac{OD} pairs, etc.

The following asymptotic approximation result provides the heavy lifting for the proof of \cref{thm:multi-variable}:

\begin{lemma}
\label{lem:benchmark}
Consider a network with nondecreasing cost functions $\test_{\edge}$,  with $\test_{\edge}(0)=0$ for $\edge\in\edges$, 
and suppose that it admits a benchmark function $\test$ at $\pole$, which is  $\tdegr$-regularly varying with $\tdegr>0$.
Consider also a sequence of inflow vectors $\rate_{\run} = \totrate_{\run} \relrate_{\run}$ such that:
\begin{enumerate}
[\indent\itshape a\hspace*{1pt}\upshape)]

\item
\label{itm:benchmark-a}
$\totrate_{\run}\to\pole$ and the vector of relative inflows $\relrate_{\run}$ converges to some $\relrate\in\simplex(\pairs)$.

\item
\label{itm:benchmark-b}
Every \ac{OD} pair has a path which is not slow \textpar{relative to $\test$}.

\item
\label{itm:benchmark-c}
There exists an \ac{OD} pair $\pair\in\pairs$ which is tight \textpar{relative to $\test$} and has $\relrate^{\pair}>0$.
\end{enumerate}
Then, the optimal allocation problem
\begin{equation}
\label{eq:tobj}
\tobj_{\run}
	= \min_{\unitflow\in\unitflows} \sum_{\edge\in\edges} \test_{\edge} \parens{\load_{\edge}(\unitflow,\rate_{\run})}
\end{equation}
satisfies
\begin{equation}
\label{eq:tobj-rate}
\lim_{\run\to\infty}\frac{\tobj_{\run}}{\test(\totrate_{\run})}=
 \vobj_{\tdegr}(\relrate),
\end{equation}
where $\vobj_{\tdegr}(\relrate)\in(0,\infty)$ is the solution value of the problem
\begin{equation}
\label{eq:vobj}
\vobj_{\tdegr} (\relrate)
	=\min_{\unitflow\in\unitflows} \sum_{\edge\in\edges} \ind_{\edge}\,\unitfunct_{\edge}(\unitflow,\relrate)^{\tdegr}
\end{equation}
and, by convention, we have set $\ind_{\edge} \unitload_{\edge}^\tdegr=0$ if $\ind_{\edge}=\infty$ and $\unitload_{\edge}=0$.
Moreover, if $\olim\unitflow_{\run}$ is a sequence of optimal solutions of $\tobj_{\run}$, every limit point of $\olim\unitflow_{\run}$ solves $\vobj_{\tdegr}(\relrate)$.
\end{lemma}

\begin{Proof}
The arguments in the proof are similar to the line of reasoning in epi-convergence arguments as in \cite{Att:Pitman1984}.
To streamline the presentation, we break up the proof in five steps as follows:

\medskip
\paragraph{\itshape Step 1:\; $\vobj_{\tdegr}(\relrate) < \infty$\afterhead}
\quad
By \condref{itm:benchmark-b}, each \ac{OD} pair admits a path that is not slow;
therefore, routing all traffic through said path gives a finite value for the objective of \eqref{eq:vobj}, implying in turn that $\vobj_{\tdegr}(\relrate)<\infty$.
More precisely, for every $\pair\in\pairs$, take a traffic allocation $\unitflow^{\pair}\in\unitflows^{\pair}$ that assigns zero weight to the slow paths $\slow\routes^{\pair}$ of $\pair$.
Then, for every slow edge $\edge\in\slow\edges$, we have $\unitload_{\edge}^{\pair}(\unitflow) = 0$ and, {\em a fortiori}, $\unitfunct_{\edge}(\unitflow,\relrate)=0$;
hence
\begin{equation}
\vobj_{\tdegr}(\relrate)
	\leq \sum_{\edge: \ind_{\edge}<\infty}
	\ind_{\edge} \,\unitfunct_{\edge}(\unitflow,\relrate)^{\tdegr}
	< \infty.
\end{equation}

\medskip
\paragraph{\itshape Step 2:\; $\vobj_{\tdegr}(\relrate) > 0$\afterhead}
\quad
By \condref{itm:benchmark-c}, there exists a tight \ac{OD} pair $\pair\in\tight\pairs$ such that $\relrate^{\pair}>0$.
For every $\unitflow\in \unitflows$ we have $\sum_{\route\in\routes^{\pair}} \unitflow_{\route} = 1$, so there exists some route $\route\in\routes^{\pair}$ with $\unitflow_{\route}^{\pair} \geq 1/\abs{\routes^{\pair}}$.
This gives $\unitload_{\edge}^{\pair}(\unitflow) \geq 1/\abs{\routes^{\pair}}$ for all $\edge\in\route$, and hence
\begin{equation}
\sum_{\edge\in\edges} \ind_{\edge} \,\unitfunct_{\edge}(\unitflow,\relrate)^{\tdegr}
	\geq \sum_{\edge\in\edges} \ind_{\edge} \parens*{\relrate^{\pair} \unitload_{\edge}^{\pair}(\unitflow)}^{\tdegr}
	\geq \sum_{\edge\in\route} \ind_{\edge} \parens*{\relrate^{\pair} / \abs{\routes^{\pair}}}^{\tdegr}
	\geq \ind^{\pair} \parens*{\relrate^{\pair} / \abs{\routes^{\pair}}}^{\tdegr}
	> 0.
\end{equation}
Minimizing over $\unitflow\in\unitflows$ then yields $\vobj_{\tdegr}(\relrate)>0$, as claimed.

\medskip
\paragraph{\itshape Step 3:\; $\limsup_{\run\to\infty} \tobj_{\run}/\test(\totrate_{\run}) \leq \vobj_{\tdegr}(\relrate)$\afterhead}
\quad Fix an optimal solution $\olim\unitflow\in\unitflows$ of \eqref{eq:vobj}.
By the finiteness of $\vobj_{\tdegr}(\relrate)$, we have 
$\unitfunct_{\edge}(\olim\unitflow,\relrate)=0$ for every slow edge $\edge\in\slow\edges$ (\ie when $\ind_{\edge}=\infty$).
If $\relrate^{\pair}>0$, this implies that $\unitload_{\edge}^{\pair}(\olim\unitflow) = 0$.
Otherwise, if $\relrate^{\pair} = 0$, the objective function of \eqref{eq:vobj} does not depend on $\unitflow^{\pair}$, so every $\unitflow^{\pair}$ with $\unitload_{\edge}^{\pair}(\unitflow) = 0$ is also optimal.
Hence we can choose the solution $\olim\unitflow$ of \eqref{eq:vobj} so that all traffic is routed along edges that are not slow.

Now, from optimality we have 
\begin{equation}
\frac{\tobj_{\run}}{\test(\totrate_{\run})}
	\leq \frac{1}{\test(\totrate_{\run})}\sum_{\edge\in\edges} {\test_{\edge}\parens*{\totrate_{\run} \unitfunct_{\edge}(\olim\unitflow,\relrate_{\run})}}.
\end{equation}
Using \cref{lem:reg-var-frac}, for every non-slow edge $\edge\in\edges\setminus\slow\edges$ (\ie $\ind_{\edge}<\infty$), we get
\begin{equation}
\label{eq:lim-edge-sup}
\lim_{\run\to\infty}\frac{\test_{\edge}(\totrate_{\run} \unitfunct_{\edge}(\olim\unitflow,\relrate_{\run}))}{\test(\totrate_{\run})}= 
	 \ind_{\edge}\, \unitfunct_{\edge}(\olim\unitflow,\relrate)^{\tdegr}.
\end{equation}
Otherwise, if $\edge\in\slow\edges$ is slow (\ie $\ind_{\edge} = \infty$), we have $\unitfunct_{\edge}(\olim\unitflow,\relrate_{\run})=0$;
thus, since $\test_{\edge}(0)=0$, we get
\begin{equation}
\lim_{\run\to\infty}\frac{\test_{\edge}(\totrate_{\run}\unitfunct_{\edge}(\olim\unitflow,\relrate_{\run}))}{\test(\totrate_{\run})}
	= \lim_{\run\to\infty}\frac{\test_{\edge}(0)}{\test(\totrate_{\run})}
	= 0
	= \ind_{\edge}\,\unitfunct_{\edge}(\olim\unitflow,\relrate)^{\tdegr}.
\end{equation}
Combining the previous three displayed equations, we obtain
\begin{equation}
\limsup_{\run\to\infty} \frac{\tobj_{\run}}{\test(\totrate_{\run})}
	\leq \sum_{\edge\in\edges} \lim_{\run\to\infty} \frac{\test_{\edge}(\totrate_{\run} \unitfunct_{\edge}(\olim\unitflow,\relrate_{\run}))}{\test(\totrate_{\run})}
	= \sum_{\edge\in\edges} \ind_{\edge}\,\unitfunct_{\edge}(\olim\unitflow,\relrate)^{\tdegr}
	= \vobj_{\tdegr}(\relrate).
\end{equation}

\medskip
\paragraph{\itshape Step 4:\; $\liminf_{\run\to\infty} \tobj_{\run}/\test(\totrate_{\run}) \geq \vobj_{\tdegr}(\relrate)$\afterhead}
\quad
Passing to a subsequence if necessary, we may assume that $\liminf_{\run\to\infty} \tobj_{\run}/\test(\totrate_{\run})$ is attained as a limit.
Thus, letting $\olim\unitflow_{\run}$ be a sequence of solutions of $\tobj_{\run}$, and taking a further subsequence if necessary, we may assume that $\olim\unitflow_{\run}$ converges to some $\olim\unitflow\in\unitflows$.
Then, ignoring the network's slow edges, we have
\begin{equation}
\frac{\tobj_{\run}}{\test(\totrate_{\run})}
	\geq \sum_{\edge:\ind_{\edge}<\infty} \frac{\test_{\edge}(\totrate_{\run}\unitfunct_{\edge}(\olim\unitflow_{\run},\relrate_{\run}))}{\test(\totrate_{\run})},
\end{equation}
and hence, by \cref{lem:reg-var-frac}, we obtain
\begin{equation}
\label{eq:lim-edge-inf}
\liminf_{\run\to\infty} \frac{\tobj_{\run}}{\test(\totrate_{\run})}
	\geq \sum_{\edge:\ind_{\edge}<\infty} \ind_{\edge}\, \unitfunct_{\edge}(\olim\unitflow,\relrate)^{\tdegr}.
\end{equation}

To proceed, we will show that $\unitfunct_{\edge}(\olim\unitflow,\relrate) = 0$ for every slow edge.
Indeed, if this were not the case, we could find some $\eps>0$ such that $\unitfunct_{\edge}(\olim\unitflow_{\run},\relrate_{\run}) > \eps$ for all sufficiently large $\run$.
With $\test_{\edge}$  nondecreasing, we then get
\begin{equation}
\frac{\tobj_{\run}}{\test(\totrate_{\run})}
	\geq \frac{\test_{\edge}(\totrate_{\run} \unitfunct_{\edge}(\olim\unitflow_{\run},\relrate_{\run}))}{\test(\totrate_{\run})}
	\geq \frac{\test_{\edge}(\totrate_{\run} \eps)}{\test(\totrate_{\run})}
	= \frac{\test_{\edge}(\totrate_{\run} \eps)}{\test(\totrate_{\run}\eps)} \frac{\test(\totrate_{\run}\eps)}{\test(\totrate_{\run})}
	\to \ind_{\edge} \eps^{\tdegr}
	= \infty,
\end{equation}
in contradiction to Steps 1 and 3 above.
From all this, it follows that
\begin{equation}
\liminf_{\run\to\infty} \frac{\tobj_{\run}}{\test(\totrate_{\run})}
	\geq \sum_{\edge\in\edges} \ind_{\edge} \,\unitfunct_{\edge}(\olim\unitflow,\relrate)^{\tdegr}
		\geq \vobj_{\tdegr}(\relrate).
\end{equation}

\medskip
\paragraph{\itshape Step 5:\; Optimality of limit points\afterhead}
\quad
As above, let $\olim\unitflow_{\run}$ be a sequence of optimal solutions of \eqref{eq:tobj} and, by descending to a subsequence if necessary, assume that it converges to some $\olim\unitflow\in\unitflows$. 
From the previous steps we have $\tobj_{\run}/\test(\totrate_{\run}) \to \vobj_{\tdegr}(\relrate)$ so, proceeding as in Step 4, we get
\begin{equation}
\vobj_{\tdegr}(\relrate)
	=\lim_{\run\to\infty} \frac{\tobj_{\run}}{\test(\totrate_{\run})}
	\geq \sum_{\edge\in\edges} \ind_{\edge}\,\unitfunct_{\edge}(\olim\unitflow,\relrate)^{\tdegr}
	\geq \vobj_{\tdegr}(\relrate),
\end{equation}
showing that $\olim\unitflow$ solves \eqref{eq:vobj}.
\end{Proof}

Armed with \cref{lem:benchmark}, we are finally in a position to prove \cref{thm:multi-variable}:
\vspace{1ex}

\begin{Proof}[Proof of \cref{thm:multi-variable}]
To begin with, express the objective function of \eqref{eq:SO} in terms of the normalized flow variables $\unitflow$ as
\begin{equation}
\label{eq:obj-inflow}
\obj_{\run}(\unitflow)
	= \sum_{\edge\in\edges} \load_{\edge}(\unitflow,\rate_{\run}) \: \cost_{\edge}(\load_{\edge}(\unitflow,\rate_{\run})).
\end{equation}
Now, let  $\eq{\unitflow}_{\run}$, $\opt{\unitflow}_{\run}$ be  the normalized traffic allocation profiles of a \acl{WE} and a \acl{SO} flow, respectively.
Then, the network's \acl{PoA} may be expressed as
\begin{equation}
\PoA(\game_{\run})
	= \frac{\Eq(\game_{\run})}{\Opt(\game_{\run})}
	= \frac{\obj_{\run}(\eq{\unitflow}_{\run})}{\obj_{\run}(\opt{\unitflow}_{\run})}.
\end{equation}
Notice that $\Opt(\game_{\run})>0$ thanks to Assumptions~\eqref{cond:nonslow} and \eqref{cond:salient}.

In order to prove that $\PoA(\game_{\run}) \to 1$ it suffices to take a subsequence $\game_{\run_k}$ realizing the 
$\limsup_{\run\to\infty}\PoA(\game_{\run})$ as a limit and to prove that $\PoA(\game_{\run_k})\to 1$.
Relabeling indices and extracting a further subsequence if necessary, we may assume without loss of generality that:
\begin{inparaenum}
[(\itshape a\upshape)]
\item the limit $\lim_{\run\to\infty} \PoA(\game_{\run})$ exists;
\item
the sequence $\relrate_{\run}$ of relative inflows converges to some $\relrate\in\simplex(\pairs)$;
and
\item
the sequences $\eq\unitflow_{\run}$ and $\opt\unitflow_{\run}$ converge to some $\eq\unitflow,\opt\unitflow\in\unitflows$ respectively.
\end{inparaenum}
With all this, we will use \cref{lem:benchmark} to derive the asymptotic behavior of $\Opt(\game_{\run})$ and $\Eq(\game_{\run})$.

First, for $\Opt(\game_{\run})$, combining \cref{lem:Karamata} with Proposition 1.5.1 of \cite{BinGolTeu:CUP1989} and the fact that the network's cost functions are nondecreasing, we immediately see that the network's benchmark function $\cost$ is $\beta$-regularly varying for some $\beta\geq0$.
Then, letting $\test_{\edge}(\load) = \load \cost_{\edge}(\load)$ and $\test(\load) = \load \cost(\load)$, we also get that $\test$ is $\tdegr$-regularly varying with $\tdegr = 1 +\beta > 0$ and $\lim_{\load\to\pole} \test_{\edge}(\load)/\test(\load) = \lim_{\load\to\pole} \cost_{\edge}(\load)/\cost(\load) = \ind_{\edge}$.
This means that the hypotheses of \cref{lem:benchmark} are all satisfied, implying in turn that
\begin{equation}
\label{eq:opt-asym}
\obj_{\run}(\opt\unitflow_{\run})
	= \Opt(\game_{\run})
	\sim \vobj_{\tdegr}(\relrate) \, \test(\totrate_{\run})
	\quad
	\text{as $\run\to\infty$},
\end{equation}
with the notation ``$f_{\run} \sim \test_{\run}$'' meaning here that $\lim_{\run\to\infty} f_{\run}/\test_{\run} = 1$.

In view of this, and since $0 < \vobj_{\tdegr}(\relrate) < \infty$, it remains to show that $\obj_{\run}(\eq{\unitflow}_{\run}) \sim \vobj_{\tdegr}(\relrate) \test(\totrate_{\run})$.
To that end, we first analyze the asymptotic behavior of the convex minimization problem
\begin{equation}
\label{eq:val}
\Val(\game_{\run})
	= \min_{\unitflow\in\unitflows} \sum_{\edge\in\edges} \Cost_{\edge}(\load_{\edge}(\unitflow,\rate_{\run}))
\end{equation}
by applying \cref{lem:benchmark} to the primitives $\Cost_{\edge}$ and $\Cost$ of $\cost_{\edge}$ and $\cost$ respectively.
By a standard result \cite[Theorem 1.5.11]{BinGolTeu:CUP1989}, $\Cost$ is $\tdegr$-regularly varying with $\tdegr = 1 + \beta$;
moreover, by L'Hôpital's rule we also have $\lim_{\load\to\pole} \Cost_{\edge}(\load)/\Cost(\load) = \lim_{\load\to\pole} \cost_{\edge}(\load)/\cost(\load) = \ind_{\edge}$.
By \cref{lem:benchmark}, it follows that $\Val(\game_{\run}) / \Cost(\totrate_{\run})\to \vobj_{\tdegr}(\relrate)$.
In addition, since the \acl{WE} traffic allocations $\eq\unitflow_{\run}$ are solutions of $\Val(\game_{\run})$, the limit $\eq\unitflow$ of $\eq\unitflow_{\run}$ 
is optimal for $\vobj_{\tdegr}(\relrate)$ by \cref{lem:benchmark}.

Noting that $\load_{\edge}(\eq\unitflow_{\run},\rate_{\run}) = \totrate_{\run} \unitfunct_{\edge}(\eq\unitflow_{\run},\relrate_{\run})$, we obtain
\begin{equation}
\label{eq:additional}
\frac{\obj_{\run}(\eq{\unitflow}_{\run})}{\test(\totrate_{\run})}
	= \sum_{\edge\in\edges}
	\frac{\test_{\edge} \parens[\big]{\totrate_{\run}\unitfunct_{\edge}(\eq\unitflow_{\run},\relrate_{\run})}}{\test(\totrate_{\run})}.
\end{equation}
By \cref{lem:reg-var-frac}, we also have the following limit for every non-slow edge $\edge\in\edges\setminus\slow\edges$:
\begin{equation}
\label{eq:comp-nonslow}
\frac{\test_{\edge}\parens*{\totrate_{\run} \unitfunct_{\edge}(\eq\unitflow_{\run},\relrate_{\run})}}{\test(\totrate_{\run})}
	\to \ind_{\edge} \, \unitfunct_{\edge}(\eq\unitflow,\relrate)^{\tdegr}.
\end{equation}
To establish a similar limiting result when $\edge$ is slow, we first claim that
there exists a constant $B\geq 0$ such that
\begin{equation}
\test_{\edge}\parens*{\totrate_{\run} \unitfunct_{\edge}(\eq\unitflow_{\run},\relrate_{\run})}
	\leq B\, \unitfunct_{\edge}(\eq\unitflow_{\run},\relrate_{\run}) \test(\totrate_{\run})
\end{equation}
This is trivially satisfied when $\unitfunct_{\edge}(\eq\unitflow_{\run},\relrate_{\run}) = 0$, so it suffices to consider the case 
$\unitfunct_{\edge}(\eq\unitflow_{\run},\relrate_{\run}) > 0$.
The above inequality is then equivalent to asking that
\begin{equation}
\label{eq:comp-slow-bound}
\cost_{\edge}\parens*{\totrate_{\run} \unitfunct_{\edge}(\eq\unitflow_{\run},\relrate_{\run})}
	\leq B \, \cost(\totrate_{\run})
\end{equation}
Now,  $\unitfunct_{\edge}(\eq\unitflow_{\run},\relrate_{\run}) > 0$ implies that the edge $\edge$ receives some equilibrium traffic from at least one \ac{OD} pair $\pair\in\pairs$, so it must  belong to a path $\route\in\routes^{\pair}$ with minimal cost.
Thus, if we consider an  alternative path $\routealt\in\routes^{\pair}$ all of whose edges are tight or fast, we have
\begin{equation}
\cost_{\edge}\parens{\totrate_{\run} \unitfunct_{\edge}(\eq\unitflow_{\run},\relrate_{\run})}
	\leq \sum_{\edgealt\in\route} \cost_{\edgealt}\parens{\totrate_{\run} \unitfunct_{\edgealt}(\eq\unitflow_{\run},\relrate_{\run})}
	\leq \sum_{\edgealt\in\routealt} \cost_{\edgealt}\parens{\totrate_{\run} \unitfunct_{\edgealt}(\eq\unitflow_{\run},\relrate_{\run})}.
\end{equation}
Using the trivial bound $\totrate_{\run} \unitfunct_{\edgealt}(\eq\unitflow_{\run},\relrate_{\run}) \leq\totrate_{\run}$, we further get 
\begin{equation}
\cost_{\edge}\parens{\totrate_{\run} \unitfunct_{\edge}(\eq\unitflow_{\run},\relrate_{\run})}
	\leq \sum_{\edgealt\in\routealt} \cost_{\edgealt}(\totrate_{\run})
	\leq \sum_{\edgealt: \ind_{\edgealt}<\infty} \cost_{\edgealt}(\totrate_{\run}).
\end{equation}
However, for every non-slow edge $\edgealt\in\edges\setminus\slow\edges$, the sequence $\cost_{\edgealt}(\totrate_{\run})/\cost(\totrate_{\run})$ converges to $\ind_{\edgealt}$
so we can find a constant $B_{\edgealt}$ such that $\cost_{\edgealt}(\totrate_{\run})/\cost(\totrate_{\run}) \leq B_{\edgealt}$ for all $n\in\N$;
consequently, \eqref{eq:comp-slow-bound} follows by taking $B=\sum_{\edgealt: \ind_{\edgealt}<\infty} B_{\edgealt}$.
Thus, given that $\eq{\unitflow}$ is optimal for $\vobj_{\tdegr}(\relrate)$,
we get $\unitfunct_{\edge}(\eq\unitflow_{\run},\relrate_{\run}) \to \unitfunct_{\edge}(\eq\unitflow,\relrate) = 0$, and hence
\begin{equation}
\label{eq:comp-slow}
\frac{\test_{\edge}\parens{\totrate_{\run} \unitfunct_{\edge}(\eq\unitflow_{\run},\relrate_{\run})}}{\test(\totrate_{\run})}
	\leq B\,\unitfunct_{\edge}(\eq\unitflow_{\run},\relrate_{\run})
	\to 0
	= \ind_{\edge} \,\unitfunct_{\edge}(\eq\unitflow,\relrate)^{\tdegr}.
\end{equation}
Combining \eqref{eq:comp-nonslow}, \eqref{eq:comp-slow} and \eqref{eq:additional}, we then get
\begin{equation}
\obj_{\run}(\eq{\unitflow}_{\run}) / \test(\totrate_{\run})
	\to \sum_{\edge\in\edges} \ind_{\edge} \,\unitfunct_{\edge}(\eq\unitflow,\relrate)^{\tdegr}
	= \vobj_{\tdegr}(\relrate),
\end{equation}
as was to be shown.
\end{Proof}

\section{Speed of convergence}
\label{app:rate}

In this appendix, we provide the proofs of the results presented in \cref{sec:rate}.


\subsection{Rates in the light traffic regime}

First we present  the proof of \cref{prop:Pigou-light} on the light traffic rates in the case of a Pigou network.

\begin{Proof}[Proof of \cref{prop:Pigou-light}]
Let $\load$ denote the flow on edge $\edge_{1}$.
At equilibrium, the costs on both edges must be equal so that $(\eq\load)^{\degr_1}=(\totrate-\eq\load)^{\degr_2}$, which is equivalent to 
$\eq\load+(\eq\load)^{\degr_1/\degr_2}=\totrate$. Since $M$ tends to 0 it follows that $\eq\load$ will be small and since
$\degr_1<\degr_2$ the  term $(\eq\load)^{\degr_1/\degr_2}$ dominates the right hand side. Therefore
\begin{equation}
\eq\load
	= \totrate^{\degr_2/\degr_1}(1+\smalloh(1))
\end{equation}
so the equilibrium cost $\Eq(\game_{\totrate}) = \totrate \cdot \cost_{1}(\eq\load)= \totrate \cdot \cost_{2}(\totrate-\eq\load)$ scales as
\begin{equation}
\Eq(\game_{\totrate})
	= \totrate \cdot \left[\totrate-\totrate^{\degr_2/\degr_1}(1+\smalloh(1))\right]^{\degr_2}
	= \totrate^{\degr_2+1} - \degr_2\totrate^{\degr_2+\degr_1/\degr_2}+\smalloh(\totrate^{\degr_2+\degr_1/\degr_2}).
\end{equation}

Similarly, if $\opt{\load}$ is the optimal flow on edge $\edge_{1}$, both edges have the same marginal cost 
\begin{equation}
(1+\degr_1)\opt{\load}^{\degr_1}
	= (1+\degr_2)(\totrate-\opt{\load})^{\degr_2}.
\end{equation}
Therefore, if we let
\begin{equation}
\label{eq:pigou alt light}
\pigoualtlight
	= \left(\frac{1+\degr_2}{1+\degr_1}\right)^{1/\degr_2},
\end{equation}
we get $\pigoualtlight\opt{\load}+\opt{\load}^{\degr_1/\degr_2}=\pigoualtlight\totrate$ as before, and hence
\begin{equation}
\opt{\load}
	=(\pigoualtlight\totrate)^{\degr_2/\degr_1}(1+\smalloh(1)).
\end{equation}
It follows that the optimal cost scales as
\begin{flalign}
\Opt(\game_{\totrate})
	&=\opt{\load} \cdot \cost_{1}(\opt{\load})+(\totrate-\opt{\load})\cdot\cost_{2}(\totrate-\opt{\load})
	\notag\\
	&=\left[(\pigoualtlight\totrate)^{\degr_2/\degr_1}(1+\smalloh(1))\right]^{\degr_1+1}
+\left[\totrate-(\pigoualtlight\totrate)^{\degr_2/\degr_1}(1+\smalloh(1))\right]^{\degr_2+1}\notag\\
	&=(\pigoualtlight\totrate)^{\degr_2+\degr_2/\degr_1} +\totrate^{\degr_2+1} - (\degr_2+1)\totrate^{\degr_2}(\pigoualtlight\totrate)^{\degr_2/\degr_1} +\smalloh(\totrate^{\degr_2+\degr_2/\degr_1})
	\notag\\
	&=\totrate^{\degr_2+1} - \pigoualtlight^{\degr_2/\degr_1}\left[(\degr_2+1)-\pigoualtlight^{\degr_2}\right]\totrate^{\degr_2+\degr_2/\degr_1}+\smalloh(\totrate^{\degr_2+\degr_2/\degr_1})
	\notag\\
	&=\totrate^{\degr_2+1} - (\pigouratelight + \degr_2) \totrate^{\degr_2+\degr_2/\degr_1}+\smalloh(\totrate^{\degr_2+\degr_2/\degr_1})
\end{flalign}
where the last equality follows from the identity $\pigoualtlight^{\degr_2/\degr_1}\left[(\degr_2+1)-\pigoualtlight^{\degr_2}\right]=\pigouratelight+\degr_2$.

Combining the previous expressions we get
\begin{flalign}
\PoA(\game_{\totrate})
	=\frac{\Eq(\game_{\totrate})}{\Opt(\game_{\totrate})}&=\frac{\Opt(\game_{\totrate}) + \pigouratelight\totrate^{\degr_2+\degr_2/\degr_1}+\smalloh(\totrate^{\degr_2+\degr_2/\degr_1})}{\Opt(\game_{\totrate})}
	\notag\\
	&=1+\frac{\pigouratelight\totrate^{\degr_2+\degr_2/\degr_1}+\smalloh(\totrate^{\degr_2+\degr_2/\degr_1})}{\totrate^{\degr_2+1}+\smalloh(\totrate^{\degr_2+1})}
	\notag\\
	&=1 + \pigouratelight\totrate^{\adegr}+\smalloh(\totrate^{\adegr}).
\end{flalign}

To complete our proof, it remains to show that $\pigourate>0$.
After a small rearrangement, this is equivalent to establishing the inequality
\begin{equation}
\parens*{\frac{1 + \degr_{2}}{1 + \degr_{1}}}^{1+\degr_{1}}
	= \parens*{1 + \frac{\degr_{2} - \degr_{1}}{1 + \degr_{1}}}^{1+\degr_{1}}
	> \parens*{1 +\frac{\degr_{2} - \degr_{1}}{\degr_{1}}}^{\degr_{1}}
	= \parens*{\frac{\degr_{2}}{\degr_{1}}}^{\degr_{1}},
\end{equation}
which itself follows from the fact that the function $(1+\gdegr/x)^{x}$ is increasing in $x$ whenever $\gdegr$ is positive.
\end{Proof}

Now we prove the following more general version of \cref{thm:rate-light}.
\begin{theorem}
\label{thm:rate-light-seq}
Let $\game_{\run}$ be a sequence of nonatomic routing games satisfying the assumptions of Theorem 
\ref{thm:multi-variable}, with $\totrate_\run\to 0$. Suppose further that the edge costs are polynomials as in \eqref{eq:polynomials}, 
and let $\adegr=\ord/\slow\ord-1$ with $\ord$ and $\slow\ord$ given by \eqref{eq:ord-net-light}
and \eqref{eq:gap-light} respectively.
Then, there exist non-negative constants $\firstconst\geq 0$ and $\secondconst\geq 0$ such that
\begin{equation}
\label{eq:rate-light-seq}
\PoA(\game_{\run})
	\leq 1 + \firstconst\totrate_{\run} + \secondconst\totrate_{\run}^{\adegr}
\end{equation} 
with $\secondconst=0$ whenever $\slow\edges=\varnothing$.
\end{theorem}

\begin{Proof} Let $\eq{\unitflow}_{\run},\opt{\unitflow}_{\run}\in\unitflows$ be an equilibrium and an optimum flow 
for $\game_{\run}$ with induced edge flows $\eq\load_{\edge,\run}=\totrate_{\run}\;\unitfunct_{\edge}(\eq\unitflow_{\run},\relrate_{\run})$ and $\opt\load_{\edge,\run}=\totrate_{\run}\;\unitfunct_{\edge}(\opt\unitflow_{\run},\relrate_{\run})$.
The social cost of $\eq{\unitflow}_{\run}$ can be estimated as
\begin{flalign}
\obj_{\run}(\eq{\unitflow}_{\run})
	&=\sum_{\edge\in\edges} 
	\eq\load_{\edge,\run}\;\cost_{\edge}\parens{\eq\load_{\edge,\run}}
	=\sum_{\edge\in\edges} \sum_{\pol=\ord_\edge}^{\degr_{\edge}}
	\cost_{\edge,\pol} \cdot \parens{\eq\load_{\edge,\run}}^{\pol+1}
	\notag\\
	&=\sum_{\edge\in\edges} \sum_{\pol=\ord_\edge}^{\degr_{\edge}}
	\bracks*{\frac{\ord+1}{\pol+1} + \frac{\pol-\ord}{\pol+1}}
	\cost_{\edge,\pol} \cdot \parens{\eq\load_{\edge,\run}}^{\pol+1}
	\notag\\
	&= (\ord+1) \sum_{\edge\in\edges} \Cost_{\edge}({\eq\load_{\edge,\run}})+
\sum_{\edge\in\edges}\sum_{\pol=\ord_\edge}^{\degr_{\edge}}\frac{\pol-\ord}{\pol+1}\cost_{\edge,\pol}\cdot\parens{\eq\load_{\edge,\run}}^{\pol+1}
	\notag\\
	&\le (\ord+1)\sum_{\edge\in\edges}\Cost_{\edge}({\opt\load_{\edge,\run}})+
\sum_{\edge\in\edges}\sum_{\pol=\ord+1}^{\degr_{\edge}}\frac{\pol-\ord}{\pol+1}\cost_{\edge,\pol}\cdot\parens{\eq\load_{\edge,\run}}^{\pol+1}
\label{oooo-light}
\end{flalign}
where $\Cost_{\edge}$ is the primitive of $\cost_{\edge}$ and for the last inequality we used the fact that ${\eq\load_{\edge,\run}}$ minimizes the first sum, 
and in the double sum we dropped the negative terms with $\pol\leq\ord$.
Now, the first sum in \eqref{oooo-light} can be further bounded as
 \begin{flalign}
(\ord+1)\sum_{\edge\in\edges}\Cost_{\edge}({\opt\load_{\edge,\run}})
	&= \sum_{\edge\in\edges} \sum_{\pol=\ord_\edge}^{\degr_{\edge}} \frac{\ord+1}{\pol+1}
	\cost_{\edge,\pol} \cdot \parens{\opt\load_{\edge,\run}}^{\pol+1}
	\notag\\
	&= \sum_{\edge\in\edges} \sum_{\pol=\ord_\edge}^{\degr_{\edge}}
	\cost_{\edge,\pol} \cdot \parens{\opt\load_{\edge,\run}}^{\pol+1}
	+ \sum_{\edge\in\edges} \sum_{\pol=\ord_\edge}^{\degr_{\edge}}
	\frac{\ord-\pol}{\pol+1} \cost_{\edge,\pol} \cdot \parens{\opt\load_{\edge,\run}}^{\pol+1}
	\notag\\
	&\leq \Opt(\game_{\run})
	+ \sum_{\edge\in\slow\edges} \sum_{\pol=\ord_\edge}^{\ord-1}\frac{\ord-\pol}{\pol+1}
	\cost_{\edge,\pol} \cdot \parens{\opt\load_{\edge,\run}}^{\pol+1}
	\notag
\end{flalign}
where  we used the optimality of $\opt\load_{\run}$ in the first double sum, 
and we dropped the negative terms in the second. 
Note that in the latter only the slow edges with $\ord_\edge<\ord$ are relevant. Combining these estimates we get
\begin{equation}\label{eq:ooo-light}
\obj_{\run}(\eq{\unitflow}_{\run})
	\leq 
	\Opt(\game_{\run})
	+ \sum_{\edge\in\slow\edges} \sum_{\pol=\ord_\edge}^{\ord-1}\frac{\ord-\pol}{\pol+1}
	\cost_{\edge,\pol} \cdot \parens{\opt\load_{\edge,\run}}^{\pol+1}
	+
\sum_{\edge\in\edges}\sum_{\pol=\ord+1}^{\degr_{\edge}}\frac{\pol-\ord}{\pol+1}\cost_{\edge,\pol}\cdot\parens{\eq\load_{\edge,\run}}^{\pol+1}
	\end{equation}

Let us call $\obj_{\run}^{\textup{I}}$ the first double sum  in \eqref{eq:ooo-light} and $\obj_{\run}^{\textup{II}}$ the second.
In order to bound $\obj_{\run}^{\textup{II}}$ we assume that $\run$  is large enough so that 
$\totrate_\run\leq 1$. This assumption is done for convenience and it only affects the value of the constants $\firstconst$ and $\secondconst$:
by redefining them appropriately, the bound \eqref{eq:rate-light-seq} will hold for all $\run$. Then  by setting 
\begin{equation}
\sumc =\sum_{\edge\in\edges} \sum_{\pol=\ord+1}^{\degr_\edge} \frac{\pol-\ord}{\pol+1} \cost_{\edge,\pol},
\end{equation}
and noting that $\eq\load_{\edge,\run}\leq\totrate_\run$, we get
\begin{equation}\label{eq:obj-II-light}
\obj_{\run}^{\textup{II}}\leq\sum_{\edge\in\edges}\sum_{\pol=\ord+1}^{\degr_{\edge}}\frac{\pol-\ord}{\pol+1}\cost_{\edge,\pol}\cdot\totrate_{\run}^{\pol+1}\leq \sumc\totrate_{\run}^{\ord+2}.
\end{equation}

In order to bound $\obj_{\run}^{\textup{I}}$ we note that this term vanishes when $\slow\edges$ is empty.
Otherwise, consider any edge $\edge\in\slow\edges$ that contributes to the sum with $\opt\load_{\edge,\run}>0$. 
We note that the optimum flow $\opt{\unitflow}_{\run}$ is an equilibrium for the marginal cost functions 
\begin{equation}\label{}
\opt\cost_\edge(\load)=\cost_\edge(\load)+\load\;\cost_{\edge}'(\load)=\sum_{\pol=\ord_\edge}^{\degr_\edge}(\pol+1)\cost_{\edge,\pol}\load^\pol.
\end{equation}
The edge $\edge$ must therefore belong to an optimal path $\route\in\routes^{\pair}$ (w.r.t. the costs 
$\opt\cost_\edge(\load)$) for some $\pair\in\pairs$.
Hence, taking any alternative path $\routealt\in\routes^{\pair}$ which is not slow, \ie with $\ord_{\edgealt}\geq\ord$ 
for all $\edgealt\in\routealt$, and denoting 
\begin{equation}
\sumcalt
	=\sum_{\edgealt\not\in\slow\edges}\sum_{\pol=\ord_{\edgealt}}^{\degr_{\edgealt}}(\pol+1)\cost_{\edgealt,\pol},
\end{equation}
we get the bound (recall that $\totrate_\run\leq 1$)
\begin{equation}\label{eq:eee-light}
\opt\cost_{\edge}(\opt\load_{\edge,\run}) 
	\leq \sum_{\edgealt\in\route}\opt\cost_{\edgealt}(\opt\load_{\edge,\run})
	\leq \sum_{\edgealt\in\routealt}\opt\cost_{\edgealt}(\opt\load_{\edge,\run})
	\leq \sum_{\edgealt\in\routealt}\opt\cost_{\edgealt}(\totrate_{\run})
	\leq \sumcalt\totrate_{\run}^{\ord}.
\end{equation}
In particular, letting $\opt\cost_{0}=\min_{\edge\in\slow\edges}(\ord_\edge+1)\cost_{\edge,\ord_{\edge}}$ we have
\begin{equation}
\opt\cost_{0}\cdot\left(\opt\load_{\edge,\run}\right)^{\ord_{\edge}}\leq
(\ord_\edge+1)\cost_{\edge,\ord_{\edge}}\cdot\left(\opt\load_{\edge,\run}\right)^{\ord_{\edge}}
	\leq \opt\cost_{\edge}\parens{\opt\load_{\edge,\run}}
	\leq \sumcalt\totrate_{\run}^{\ord}.
\end{equation}
Now, for $\run$ large we have $\sumcalt\totrate_{\run}^{\ord}/\opt\cost_0\leq 1$  
and since $\ord_\edge\leq\slow\ord$ we obtain
$\opt\load_{\edge,\run}\le\left(\sumcalt\totrate_{\run}^{\ord}/\opt\cost_{0}\right)^{1/\slow\ord}$. 
Combining this latter bound with \eqref{eq:eee-light},  and
denoting $\sumB = \sumcalt\left(\sumcalt/\opt\cost_{0}\right)^{1/\slow\ord}|\slow\edges|$, we deduce
\begin{flalign}\label{eq:obj-I-light}
\obj_{\run}^{\textup{I}} &=
\sum_{\edge\in\slow\edges} \sum_{\pol=\ord_\edge}^{\ord-1}\frac{\ord-\pol}{\pol+1}
	\cost_{\edge,\pol} \cdot \parens{\opt\load_{\edge,\run}}^{\pol+1}\nonumber \\
	&\leq
\sum_{\edge\in\slow\edges} \sum_{\pol=\ord_\edge}^{\ord-1}(\pol+1)	\cost_{\edge,\pol} \cdot \parens{\opt\load_{\edge,\run}}^{\pol+1}\nonumber \\
&\leq
\sum_{\edge\in\slow\edges} \opt\load_{\edge,\run}\;\opt\cost_{\edge}(\opt\load_{\edge,\run})\nonumber \\
	&\leq \sum_{\edge\in\slow\edges} \parens*{\sumcalt\totrate_{\run}^{\ord}/\opt\cost_{0}}^{1/\slow\ord}\sumcalt\totrate_{\run}^{\ord} \nonumber\\
&\le\sumB\totrate_{\run}^{\ord+\ord/\slow\ord}.
\end{flalign}

Plugging \eqref{eq:obj-I-light} and \eqref{eq:obj-II-light} into \eqref{eq:ooo-light} we get
\begin{flalign}
\label{eq:poapol-light}
\PoA(\game_{\run})
	=\frac{\obj_{\run}(\eq{\unitflow}_{\run})}{\Opt(\game_{\run})}
	&\leq \frac{\Opt(\game_{\run}) + \sumc\totrate_{\run}^{\ord+2} + \sumB\totrate_{\run}^{\ord+\ord/\slow\ord}}{\Opt(\game_{\run})}.
\end{flalign}

Now, if we set $\minc=\min_{\edge\in\edges}\cost_{\edge,\ord_{\edge}}$, we have the following lower bound for the optimal cost
\begin{flalign}
\Opt(\game_{\run})
	&= \sum_{\edge\in\edges}
	\opt\load_{\edge,\run}
	\cdot \cost_{\edge}(\opt\load_{\edge,\run})
	\geq \minc \sum_{\edge\in\edges}
	\parens{\opt\load_{\edge,\run}}^{\ord_{\edge}+1}.
\end{flalign}

We claim that the latter is of order at least $O(\totrate_{\run}^{\ord+1})$. Indeed, let us take $\eps>0$ with $\sum_{\pair\in\tight\pairs} \relrate_{\run}^\pair\geq\eps$ for sufficiently large $\run$.
For each $\run\in\N$ we may find $\pair\in\tight\pairs$ such that $\relrate_{\run}^{\pair} \geq \eps/\abs{\tight\pairs}$ and, similarly, there exists a path $\route\in\routes^{\pair}$ with $\opt\unitflow_{\route,\run} \geq 1/\abs{\routes^{\pair}}\geq 1/\abs{\routes}$.
Then, setting $\kappa=1/(\abs{\tight\pairs}\times\abs{\routes})$ we have $\unitfunct_{\edge}(\opt\unitflow_\run,\relrate_{\run}) \geq \kappa\eps$ and therefore $\opt\load_{\edge,\run} \geq \totrate_\run\kappa\eps$ for all $\edge\in\route$.
For $\run$ large we may assume that $\totrate_{\run}\kappa\eps\leq 1$ and, since the path $\route$ contains at least one edge $\edge\in\route$ with $\ord_{\edge}\leq\ord$, setting $\bar{\minc}= \minc(\kappa\eps)^{\ord+1}$ we get
\begin{equation}
\Opt(\game_{\run})
	\geq \minc\left(\totrate_{\run}\kappa\eps\right)^{\ord_{\edge}+1}\ge\bar{\minc}\totrate_{\run}^{\ord+1}.
\end{equation}

This lower bound, combined with \eqref{eq:poapol-light}, yields \eqref{eq:rate-light-seq} with $\firstconst=\sumc/\bar{\minc}$ and $\secondconst=\sumB/\bar{\minc}$.
We conclude by noting that when $\slow\edges=\varnothing$ we have $\obj_{\run}^{\textup{I}} = 0$ and therefore we may take $\secondconst=0$.
\end{Proof}


\subsection{Rates in the heavy traffic regime}


We proceed with the proof of \cref{prop:Pigou-heavy} on the heavy traffic rates in the case of a Pigou network.
\begin{Proof}[Proof of \cref{prop:Pigou-heavy}]
Let $\load$ denote the flow on edge $\edge_{2}$.
At equilibrium, the costs on both edges must be equal so $(\eq\load)^{\degr_{2}}=(\totrate-\eq\load)^{\degr_{1}}$, which is equivalent to  $\eq\load+(\eq\load)^{\degr_{2}/\degr_{1}} = \totrate$.
It thus follows that
\begin{equation}
\eq\load
	= \totrate^{\degr_{1}/\degr_{2}}(1+\smalloh(1)),
\end{equation}
implying in turn that the equilibrium cost $\Eq(\game_{\totrate}) = \totrate \cdot \cost_{2}(\eq\load)= \totrate \cdot \cost_{1}(\totrate-\eq\load)$ scales as
\begin{equation}
\Eq(\game_{\totrate})
	= \totrate \cdot \left[\totrate-\totrate^{\degr_{1}/\degr_{2}}(1+\smalloh(1))\right]^{\degr_{1}}
	= \totrate^{1+\degr_{1}} - \degr_{1}\totrate^{\degr_{1} + \degr_{1}/\degr_{2}}+\smalloh(\totrate^{\degr_{1} + \degr_{1}/\degr_{2}}).
\end{equation}
Similarly, if $\opt{\load}$ is the optimal flow on edge $\edge_{2}$, both edges have the same marginal cost, namely
\begin{equation}
(1+\degr_{2})\opt{\load}^{\degr_{2}}
	= (1+\degr_{1})(\totrate-\opt{\load})^{\degr_{1}},
\end{equation}
and hence
\begin{equation}
\opt{\load}
	=(\pigoualt\totrate)^{\degr_{1}/\degr_{2}}(1+\smalloh(1)),
\end{equation}
where we set
\begin{equation}
\label{eq:pigou alt}
\pigoualt
	= \left(\frac{1+\degr_{1}}{1+\degr_{2}}\right)^{1/\degr_{1}}.
\end{equation}
Therefore, the optimal social cost scales as
\begin{flalign}
\Opt(\game_{\totrate})
	&=(\totrate-\opt{\load})\cdot\cost_{1}(\totrate-\opt{\load})+\opt{\load} \cdot \cost_{2}(\opt{\load})
	\notag\\
	&= \bracks[\big]{\totrate-(\pigoualt\totrate)^{\degr_{1}/\degr_{2}}(1+\smalloh(1))}^{\degr_{1}+1}
	+ \bracks[\big]{(\pigoualt\totrate)^{\degr_{1}/\degr_{2}}(1+\smalloh(1))}^{\degr_{2}+1}
	\notag\\
	&=\totrate^{\degr_{1}+1} - (\degr_{1}+1)\totrate^{\degr_{1}}(\pigoualt\totrate)^{\degr_{1}/\degr_{2}}
	+ (\pigoualt\totrate)^{\degr_{1} + \degr_{1}/\degr_{2}} +\smalloh(\totrate^{\degr_{1} + \degr_{1}/\degr_{2}})
	\notag\\
&=\totrate^{\degr_{1}+1}
	- \pigoualt^{\degr_{1}/\degr_{2}} \bracks{(\degr_{1}+1) - \pigoualt^{\degr_{1}}} \totrate^{\degr_{1} + \degr_{1}/\degr_{2}}
	+\smalloh(\totrate^{\degr_{1} + \degr_{1}/\degr_{2}}),\\
	&=\totrate^{\degr_{1}+1}
	- (\pigourate + \degr_{1}) \totrate^{\degr_{1} + \degr_{1}/\degr_{2}}
	+\smalloh(\totrate^{\degr_{1} + \degr_{1}/\degr_{2}}),
\end{flalign}
where $\pigourate$ is defined as in \eqref{eq:Pigou-heavy}.

Combining the previous expressions, we then get
\begin{flalign}
\PoA(\game_{\totrate})
	&= \frac{\Opt(\game_{\totrate}) + \pigourate\totrate^{\degr_{1} + \degr_{1}/\degr_{2}} + \smalloh(\totrate^{\degr_{1} + \degr_{1}/\degr_{2}})}{\Opt(\game_{\totrate})}\\
	&= 1 + \frac{\pigourate\totrate^{\degr_{1} + \degr_{1}/\degr_{2}} + \smalloh(\totrate^{\degr_{1} + \degr_{1}/\degr_{2}})}{\totrate^{\degr_{1}+1} + \smalloh(\totrate^{\degr_{1}+1})}
	\notag\\[.5ex]
	&= 1 + \pigourate\totrate^{-\adegr} + \smalloh\parens*{\totrate^{-\adegr}},
\end{flalign}
which establishes the first part of \cref{prop:Pigou-heavy}.
Finally, the positivity of $\pigouratelight>0$, follows again from the fact that $(1-\gdegr/x)^{x}$  increased 
with $x$ when $\gdegr>0$.
\end{Proof}

The following more general result subsumes \cref{thm:rate-heavy}.

\begin{theorem}
\label{thm:rate-heavy_sequential}
Let $\game_{\run}$ be a sequence of nonatomic routing games satisfying the assumptions of Theorem 
\ref{thm:multi-variable}, with $\totrate_\run\to\infty$. Suppose further that the edge costs are polynomials as in \eqref{eq:polynomials}, 
and let $\adegr=1-\degr/\slow\degr$ with $\degr$ and $\slow\degr$ given by \eqref{eq:ord-net-heavy}
and \eqref{eq:gap-heavy}.  Then there exist non-negative constants $\firstconst\geq 0$ and $\secondconst\geq 0$ such that
\begin{equation}
\label{eq:rate-heavy-seq}
\PoA(\game_{\run})
	\leq 1 + \frac{\firstconst}{\totrate_{\run}} + \frac{\secondconst}{\totrate_{\run}^{\adegr}}
\end{equation} 
with $\secondconst=0$ whenever $\slow\edges=\varnothing$.
\end{theorem}

\begin{Proof}
The proof follows a similar pattern as the one of 
\cref{thm:rate-light-seq}.
Let again $\eq{\unitflow}_{\run},\opt{\unitflow}_{\run}\in\unitflows$ be an equilibrium and an optimum for $\game_{\run}$, respectively. Denote $\eq\load_{\edge,\run}=\totrate_{\run}\;\unitfunct_{\edge}(\eq\unitflow_{\run},\relrate_{\run})$
and $\opt\load_{\edge,\run}=\totrate_{\run}\;\unitfunct_{\edge}(\opt\unitflow_{\run},\relrate_{\run})$ the corresponding
induced edge flows.
As before, the social cost of $\eq{\unitflow}_{\run}$ can be estimated as
\begin{flalign}
\obj_{\run}(\eq{\unitflow}_{\run})
	&=\sum_{\edge\in\edges} 
	\eq\load_{\edge,\run}\;\cost_{\edge}\parens{\eq\load_{\edge,\run}}
	=\sum_{\edge\in\edges} \sum_{\pol=\ord_\edge}^{\degr_{\edge}}
	\cost_{\edge,\pol} \cdot \parens{\eq\load_{\edge,\run}}^{\pol+1}
	\notag\\
	&=\sum_{\edge\in\edges} \sum_{\pol=\ord_\edge}^{\degr_{\edge}}
	\bracks*{\frac{\degr+1}{\pol+1} + \frac{\pol-\degr}{\pol+1}}
	\cost_{\edge,\pol} \cdot \parens{\eq\load_{\edge,\run}}^{\pol+1}
	\notag\\
	&= (\degr+1) \sum_{\edge\in\edges} \Cost_{\edge}({\eq\load_{\edge,\run}})+
\sum_{\edge\in\edges}\sum_{\pol=\ord_\edge}^{\degr_{\edge}}\frac{\pol-\degr}{\pol+1}\cost_{\edge,\pol}\cdot\parens{\eq\load_{\edge,\run}}^{\pol+1}
	\notag\\
	&\le (\degr+1)\sum_{\edge\in\edges}\Cost_{\edge}({\opt\load_{\edge,\run}})+
\sum_{\edge\in\slow\edges}{\eq\load_{\edge,\run}}\cdot\cost_{\edge}({\eq\load_{\edge,\run}}),\label{oooo}
\end{flalign}
where  in the last inequality we used the fact that ${\eq\load_{\edge,\run}}$ minimizes the first sum, while in the double sum we dropped the edges $\edge\not\in\slow\edges$ since $(\pol-\degr)/(\pol+1)\le 0$ for all $\pol\leq\degr_{\edge}\leq\degr$, and we used the inequality $(\pol-\degr)/(\pol+1)\le 1$ to bound the remaining terms $\edge\in\slow\edges$ by factoring out ${\eq\load_{\edge,\run}}$ and using the expression \eqref{eq:polynomials} for $\cost_{\edge}(\load)$.
Now, the first sum in \eqref{oooo} can be further bounded as
 \begin{flalign}
(\degr+1)\sum_{\edge\in\edges}\Cost_{\edge}({\opt\load_{\edge,\run}})
	&= \sum_{\edge\in\edges} \sum_{\pol=\ord_\edge}^{\degr_{\edge}} \frac{\degr+1}{\pol+1}
	\cost_{\edge,\pol} \cdot \parens{\opt\load_{\edge,\run}}^{\pol+1}
	\notag\\
	&= \sum_{\edge\in\edges} \sum_{\pol=\ord_\edge}^{\degr_{\edge}}
	\cost_{\edge,\pol} \cdot \parens{\opt\load_{\edge,\run}}^{\pol+1}
	+ \sum_{\edge\in\edges} \sum_{\pol=\ord_\edge}^{\degr_{\edge}}
	\frac{\degr-\pol}{\pol+1} \cost_{\edge,\pol} \cdot \parens{\opt\load_{\edge,\run}}^{\pol+1}
	\notag\\
	&\leq \Opt(\game_{\run})
	+ \sum_{\edge\in\edges} \sum_{\pol=\ord_\edge}^{\degr-1}\frac{\degr-\pol}{\pol+1}
	\cost_{\edge,\pol} \cdot \parens{\opt\load_{\edge,\run}}^{\pol+1},
	\notag
\end{flalign}
where in the inequality we used the optimality of $\opt\load_{\run}$ for the first sum and we dropped the negative terms in the second sum. Putting all this together we obtain the bound
\begin{equation}\label{eq:boundLnyn}
\obj_{\run}(\eq{\unitflow}_{\run})
	\leq 
	\Opt(\game_{\run})
	+ \sum_{\edge\in\edges} \sum_{\pol=\ord_\edge}^{\degr-1}\frac{\degr-\pol}{\pol+1}
	\cost_{\edge,\pol} \cdot \parens{\opt\load_{\edge,\run}}^{\pol+1} +
\sum_{\edge\in\slow\edges}{\eq\load_{\edge,\run}}\cdot\cost_{\edge}({\eq\load_{\edge,\run}}).
	\end{equation}

Now,  call $\obj_{\run}^{\textup{I}}$ the first double sum and $\obj_{\run}^{\textup{II}}$ the last sum in \eqref{eq:boundLnyn}.
In order to bound $\obj_{\run}^{\textup{I}}$ we assume that $\run$ is large enough so that $\totrate_\run\geq 1$.
Then, denoting 
\begin{equation}
\sumc
	=\sum_{\edge\in\edges} \sum_{\pol=\ord_\edge}^{\degr-1} \frac{\degr-\pol}{\pol+1} \cost_{\edge,\pol}
\end{equation}
and using the fact that ${\opt\load_{\edge,\run}} \leq \totrate_\run$, we can bound $\obj_{\run}^{\textup{I}}$ as 
\begin{equation}
\label{eq:obj-I}
\obj_{\run}^{\textup{I}}
	\leq\sum_{\edge\in\edges} \sum_{\pol=\ord_\edge}^{\degr-1} \frac{\degr-\pol}{\pol+1}
	\cost_{\edge,\pol} \cdot \totrate_{\run}^{\pol+1}
\leq
	\sumc\totrate_{\run}^{\degr}.
\end{equation}

In order to bound $\obj_{\run}^{\textup{II}}$ we note that this term vanishes whenever $\slow\edges$ is empty.
Otherwise, consider any edge $\edge\in\slow\edges$ that contributes to the sum with $\eq\load_{\edge,\run}>0$. Since $\eq{\unitflow}_{\run}$ is an equilibrium, the edge $\edge$ must belong to a path $\route\in\routes^{\pair}$ with minimal cost for some $\pair\in\pairs$.
Hence, taking any alternative path $\routealt\in\routes^{\pair}$ which is not slow, and denoting 
\begin{equation}
\sumcalt
	=\sum_{\edgealt\not\in\slow\edges}\sum_{\pol=\ord_\edgealt}^{\degr_{\edgealt}}\cost_{\edgealt,\pol},
\end{equation}
we get the bound
\begin{flalign}
\cost_{\edge}(\eq\load_{\edge,\run}) 
	&\leq \sum_{\edgealt\in\route}\cost_{\edgealt}(\eq\load_{\edge,\run})
	\leq \sum_{\edgealt\in\routealt}\cost_{\edgealt}(\eq\load_{\edge,\run})
	\leq \sum_{\edgealt\in\routealt}\cost_{\edgealt}(\totrate_{\run})
	\leq \sumcalt\totrate_{\run}^{\degr}.\label{eee}
\end{flalign}
In particular, letting $\cost_{0}=\min_{\edge\in\slow\edges}\cost_{\edge,\degr_{\edge}}$ we have
\begin{equation}
\cost_{0}\cdot\left(\eq\load_{\edge,\run}\right)^{\degr_{\edge}}\leq
\cost_{\edge,\degr_{\edge}}\cdot\left(\eq\load_{\edge,\run}\right)^{\degr_{\edge}}
	\leq \cost_{\edge}\parens{\eq\load_{\edge,\run}}
	\leq \sumcalt\totrate_{\run}^{\degr}.
\end{equation}
Now, for $\run$ large we have $\sumcalt\totrate_{\run}^{\degr}/\cost_0\geq 1$  
and since $\degr_\edge\geq\slow\degr$ we get
$\eq\load_{\edge,\run}\le\left(\sumcalt\totrate_{\run}^{\degr}/\cost_{0}\right)^{1/\slow\degr}$. 
Combining this latter bound with \eqref{eee},  and
denoting $\sumB = \sumcalt\left(\sumcalt/\cost_{0}\right)^{1/\slow\degr}|\slow\edges|$, we deduce
\begin{equation}
\label{eq:obj-II}
\obj_{\run}^{\textup{II}} = \sum_{\edge\in\slow\edges}{\eq\load_{\edge,\run}}\cdot\cost_{\edge}({\eq\load_{\edge,\run}})
	\leq \sum_{\edge\in\slow\edges} \parens*{\sumcalt\totrate_{\run}^{\degr}/\cost_{0}}^{1/\slow\degr}\sumcalt\totrate_{\run}^{\degr}
\le\sumB\totrate_{\run}^{\degr+\degr/\slow\degr}.
\end{equation}

Plugging \eqref{eq:obj-I} and \eqref{eq:obj-II} into \eqref{eq:boundLnyn} we get
\begin{flalign}
\label{eq:poapol}
\PoA(\game_{\run})
	=\frac{\obj_{\run}(\eq{\unitflow}_{\run})}{\Opt(\game_{\run})}
	&\leq \frac{\Opt(\game_{\run}) + \sumc\totrate_{\run}^{\degr} + \sumB\totrate_{\run}^{\degr+\degr/\slow\degr}}{\Opt(\game_{\run})}.
\end{flalign}

Now, if we set $\minc=\min_{\edge\in\edges}\cost_{\edge,\degr_{\edge}}$, we have the following lower bound for the optimal cost
\begin{flalign}
\Opt(\game_{\run})
	&= \sum_{\edge\in\edges}
	\opt\load_{\edge,\run}
	\cdot \cost_{\edge}(\opt\load_{\edge,\run})
	\geq \minc \sum_{\edge\in\edges}
	\parens{\opt\load_{\edge,\run}}^{\degr_{\edge}+1}.
\end{flalign}

We claim that the latter is of order at least $O(\totrate_{\run}^{\degr+1})$. Indeed, let us take $\eps>0$ with $\sum_{\pair\in\tight\pairs} \relrate_{\run}^\pair\geq\eps$ for sufficiently large $\run$.
For each $\run\in\N$ we may find $\pair\in\tight\pairs$ such that $\relrate_{\run}^{\pair} \geq \eps/\abs{\tight\pairs}$ and, similarly, there exists a path $\route\in\routes^{\pair}$ with $\opt\unitflow_{\route,\run} \geq 1/\abs{\routes^{\pair}}\geq 1/\abs{\routes}$.
Then, setting $\kappa=1/(\abs{\tight\pairs}\times\abs{\routes})$ we have $\unitfunct_{\edge}(\opt\unitflow_\run,\relrate_{\run}) \geq \kappa\eps$ and therefore $\opt\load_{\edge,\run} \geq \totrate_\run\kappa\eps$ for all $\edge\in\route$.
For $\run$ large we may assume that $\totrate_{\run}\kappa\eps\geq 1$ and, since the path $\route$ contains at least one edge $\edge\in\route$ with $\degr_{\edge}\geq\degr$, setting $\bar{\minc }= \minc (\kappa\eps)^{\degr+1}$ we get
\begin{equation}
\Opt(\game_{\run})
	\geq \minc\left(\totrate_{\run}\kappa\eps\right)^{\degr_{\edge}+1}\ge\bar{\minc}\totrate_{\run}^{\degr+1}.
\end{equation}

This lower bound, combined with \eqref{eq:poapol}, yields \eqref{eq:rate-heavy-seq} with $\firstconst=\sumc/\bar{\minc}$ and $\secondconst=\sumB/\bar{\minc}$.
We conclude by noting that when $\slow\edges=\varnothing$ we have $\obj_{\run}^{\textup{II}} = 0$ and therefore we may take $\secondconst=0$.
\end{Proof}

\bibliographystyle{apalike}
\bibliography{../bibtex/IEEEabrv,../bibtex/biblimitpoa}

\end{document}